\documentclass[12pt,preprint]{aastex}
\begin{document}
\title{The orbits of 48 globular clusters in a Milky-Way-Like Barred Galaxy}

\author{Christine Allen\altaffilmark{1}, Edmundo
  Moreno\altaffilmark{1} and B\'arbara Pichardo\altaffilmark{1,2} }
\altaffiltext{1}{Instituto de Astronom\'\i a, Universidad Nacional
  Aut\'onoma de M\'exico, Apdo. Postal 70-264, 04510, M\'exico, D.~F.,
  M\'exico.}  \altaffiltext{2}{Institute for Theoretical Physics,
  University of Zurich, Winterthurerstrasse 190, Zurich 8057,
  Switzerland.}

\begin{abstract} 
  The effect of a barred potential (such as the one of the Milky Way)
  on the galactic orbits of forty-eight globular clusters for which
  absolute proper motions are known is studied.  The orbital
  characteristics are compared with those obtained for the case of an
  axisymmetric galactic potential. Tidal radii are computed and
  discussed for both the better known axisymmetric case and that
  including a bar. The destruction rates due to bulge and disk
  shocking are calculated and compared in both galactic potentials.

\end{abstract}

\keywords{galaxy: halo --- galaxy: kinematics and dynamics --- globular clusters:
general}

\section{INTRODUCTION}\label{introd}

A considerable number of absolute proper motions for globular clusters
have become available in recent times, largely due to the work of
\citet{D97,D99a,D99b,D00,D01,D03}. With the new proper motions, full
space velocities can now be calculated. Then, using a model for the
galactic mass distribution, it is possible to study the galactic
orbits of a good fraction -almost a third- of the known globular
clusters. Vice-versa, through their orbital motions, globular clusters
can serve as kinematical probes of the gravitational field of our
Galaxy.

Some of the first efforts in orbit computations for globular clusters
are due to the work of \citet{KIH73} and \citet{KI75} who numerically
integrated orbits for M67, NGC 188, Omega Cen, and NGC 2420 using full
space motions in a Schmidt-type potential. With a more realistic
galactic potential that included a massive halo, \citet{AM88} and
\citet{A90} computed orbits for a total of 15 clusters -all those with
then available absolute proper motions. Using an improved version of
their axisymmetric potential, \citet{AS93} recomputed these orbits and
added those of a further six clusters whose absolute proper motions
had been determined in the intervening years.

It is only quite recently that vastly improved measures for a
significant number of clusters have become available
\citep{D97,D99a,D99b,D00, D01,D03}.  These authors use the Yale proper
motion survey of the southern sky and measure motions directly with
respect to external galaxies, or in difficult cases, to Hipparcos
reference stars. \citet{D99b} have also used the full space motions
(obtained from their proper motions and radial velocities taken from
the literature) to compute galactic orbits for 38 clusters in several
models for axisymmetric potentials. From the galactic orbits they have
derived a number of interesting conclusions.

The importance of using full space motions -instead of only radial
velocities- when attempting to derive characteristics of the globular
cluster system has been repeatedly emphasized \citep{W98,D99b,K04}.
Indeed, using only partial information -radial velocities- makes it
necessary to assume some statistical distribution for two components
of the space velocity.  With the availability of full space motions
for about a third of the known clusters, this approach seems less and
less justified, especially since in some cases its results may be
misleading.

However, since evidence for our Galaxy harboring a bar has become
increasingly convincing, it appears necessary to study the influence
of this bar on the galactic motions of the clusters as obtained from
their full space velocities and employing a galactic potential that
includes a bar. These orbits can serve the purpose of calculating
tidal shock effects on the cluster.  They can also serve as
kinematical probes of the inner part of our Galaxy, and so aid in
constraining some dynamical parameters of the bar, or at least in
establishing whether or not the model potential -especially the bar-
is compatible with observed characteristics of the clusters.

In this paper, we analyze the orbits in a galactic potential with a
bar of 48 globular clusters with available absolute proper
motions. The potential we use is that of \citet{PMM04}, which is based
on the axisymmetric model of \citet{AS91} and includes a moderately
strong bar. We compare the orbits with those obtained in the
axisymmetric potential, devoting special attention to orbits that
reside entirely within the bar region.

The orbits generally do not show secular changes in the total energy,
$E$, or in the z-component of the angular momentum, $h$, both computed
in the inertial galactic frame. Indeed, with few exceptions, those
quantities are on the average conserved within 10 percent. However in
a few cases, sudden changes in $E$ and $h$ do occur. Except for orbits
residing entirely within the bar region, the orbits do not differ
considerably from those obtained in the axisymmetric case. This
similarity also applies to the tidal-shock destruction rates computed
in both galactic potentials.

In Section \ref{gpot} the galactic potential and the observational
data to compute the initial conditions of the clusters' orbits are
presented.  In Section \ref{galorb} we show meridional orbits of some
clusters, and some effects of the bar are considered.  The effect of
the bar on the estimated tidal radii is studied in Section
\ref{tirad}, and compared with the axisymmetric case.  Destruction
rates of clusters in both the axisymmetric and non-axisymmetric
potentials are studied in Section \ref{destr}. Section \ref{concl}
presents our conclusions.

\section{THE GALACTIC POTENTIAL AND THE INITIAL CONDITIONS}\label{gpot}

For this study we have employed the axisymmetric galactic potential of
\citet{AS91}, and the barred galactic potential of \citet{PMM04}. The
axisymmetric potential assembles three components, a bulge and a
flattened disk of the form proposed by \citet{MN75}, and a massive
spherical halo extending to a radius of 100 kpc. The model satisfies
the usual observational constraints such as the galactic rotation
curve, the perpendicular force at the solar neighborhood, and the
local escape velocity ($\approx$ 536 km s$^{-1}$). The adopted
parameters of the model are $R_0=8.5$ kpc, and a circular velocity at
the Sun's position $V(R_0)=220$ km s$^{-1}$. The total mass is $9
\times 10^{11} \ M_\odot$. The local total mass density is
$\rho_0=0.15 \ M_\odot$ pc$^3$. The resulting Oort constants are
$A=12.95$ km s$^{-1}$ kpc$^1$ and $B=-12.93$ km $s^{-1}$ kpc$^1$. In
this work we have used this model as the underlying axisymmetric
potential, but replacing 70\% of the bulge mass with a bar.

The potential of the bar is one of the three constructed in
\citet{PMM04}. The one we have selected represents the best fit to the
most important observational characteristics of the galactic bar. It
consists of a superposition of four ellipsoids, each one with a
self-similar mass distribution (Schmidt 1956) to simulate the total
density profile. The superposition of the ellipsoids is carried out so
as to reproduce the apparent ``boxy'' edge-on brightness profile of
the galactic bar. With the selected parameters (chosen from a
compilation of recent observations, see references in \citet{PMM04})
the bar included in our model has a major semiaxis of 3.13 kpc, axial
ratios of 1.7:0.64:0.44, and moves with an angular velocity of 60  km
s$^{-1}$ kpc$^{-1}$. In this way, the bar model closely approximates
the Model S of \citet{F98}, based on COBE/DIRBE infrared observations 
of the galactic bar. Throughout this study we will use this so-called
``superposition model'', which is the most accurate, albeit the most
complicated numerically.

In their study, \citet{PMM04} obtained galactic orbits for three
clusters, NGC 5139, NGC 6093, and NGC 6218, selected because their
orbits in axisymmetric potentials indicated that they reside mostly
within the region of the bar, and so the influence of the bar on their
orbits was expected to be large, as indeed it turned out to be. We now
extend this study to the complete sample of clusters.

For the initial conditions we take the absolute proper motions
provided by \citet{D99b,D00,D01,D03}, but we adopt for 47 Tuc (NGC
104) and M4 (NGC 6121) the values derived from HST measures
\citep{AK03,BPKA03}.  Other data are taken from the compilation by
\citet{H96}. We consider the solar motion as
$(U,V,W)_\odot=(-10,5.2,7.2)$ km s$^{-1}$ \citep{BM98}.  Once the
space velocities are obtained, we integrate the orbits backwards in
time for 1.6 $\times 10^{10}$ years. For the integration we use the
Bulirsch-Stoer algorithm of \citet{P92}. In the axisymmetric case the
relative errors in the total energy were $10^{-14}$ at the end of the
integration. In the barred potential, the orbits are computed in the
non-inertial reference frame where the bar is at rest.  In this case
the precision of the calculations can be checked using Jacobi's
constant. The relative errors in this quantity turn out to be,
typically, $10^{-10}-10^{-11}$.

\section{THE GALACTIC ORBITS}\label{galorb}

\begin{figure}
\plotone{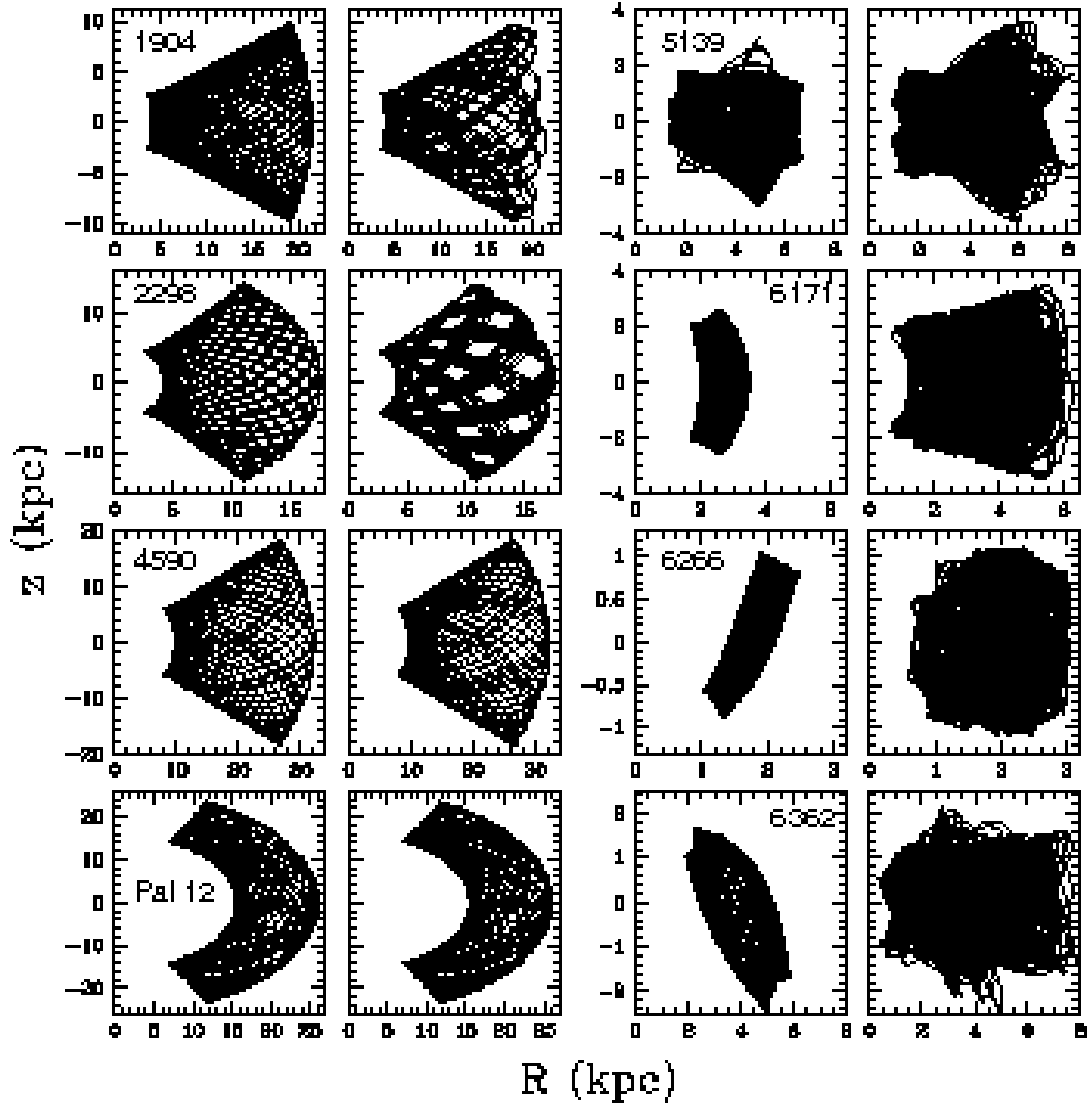}
\caption {Meridional galactic orbits of some clusters. In each pair of
frames, the orbits with the axisymmetric and non-axisymmetric
potentials are shown on the left and right, respectively. The NGC or
Pal number is written in the left frame. In general the scales are not
the same in the horizontal and vertical axes.}
\label{fig1}
\end{figure}

\begin{figure}
\plotone{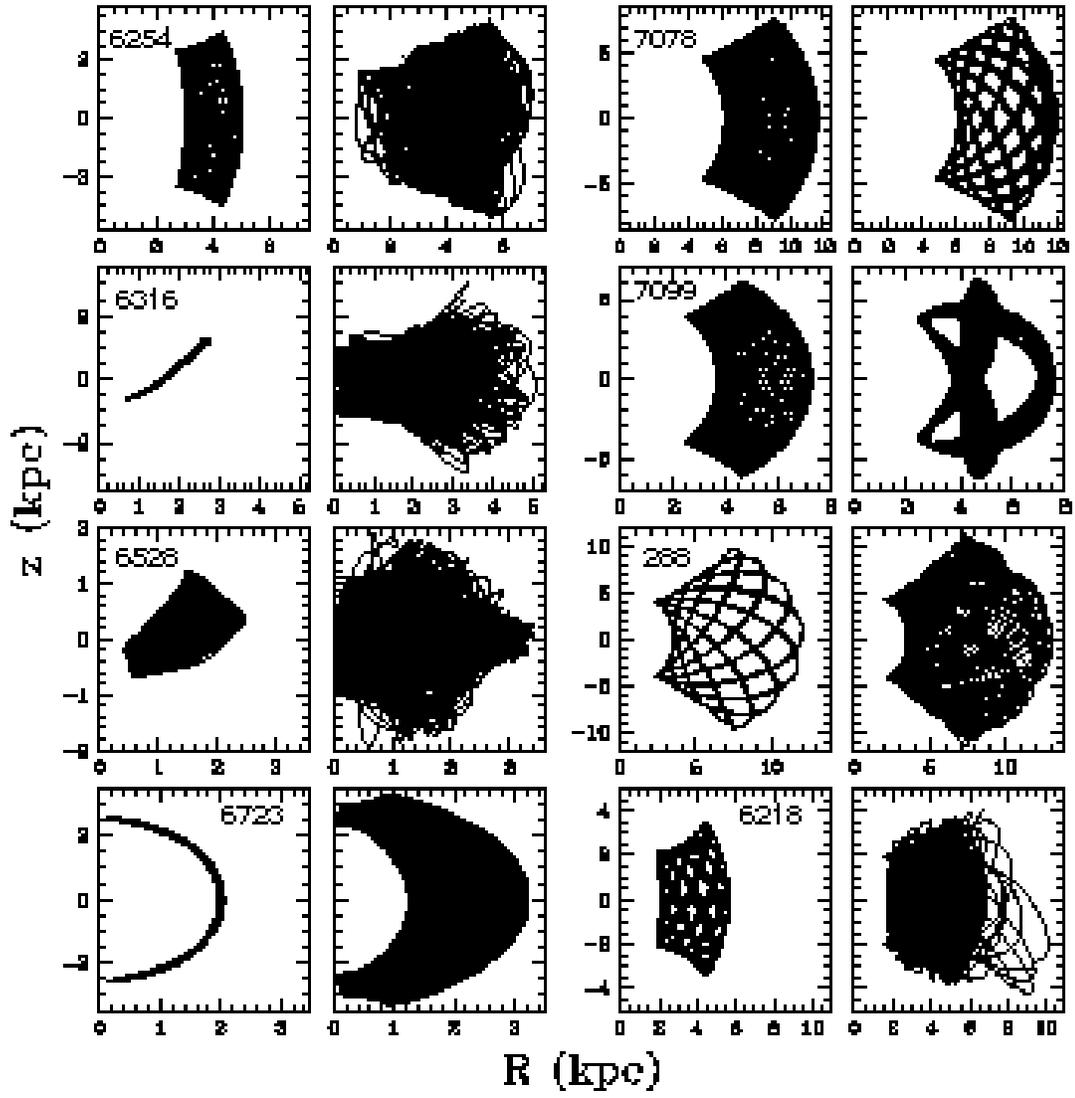}
\caption {As in Figure \ref{fig1}.}
\label{fig2}
\end{figure}

We have computed orbits for the full sample of 48 clusters. The orbits
are computed in the axisymmetric and non-axisymmetric
potentials. Figures \ref{fig1} and \ref{fig2} show a few meridional
orbits as representative examples.  For comparison, for a given
cluster we show the meridional orbits in both galactic
potentials. Tables \ref{tbl-1} and \ref{tbl-2} summarize our results
for both cases. In Table \ref{tbl-1}, corresponding to the
axisymmetric potential, successive columns contain the name of the
cluster, the minimum perigalactic distance reached in the course of
the complete orbit, the average perigalactic distance, the maximum
apogalactic distance, the average apogalactic distance, the maximum
distance from the galactic plane reached throughout the entire orbit,
the average maximum distance from the galactic plane, the average
orbital eccentricity, the orbital energy per unit mass, the z-
component of angular momentum per unit mass, two values for the
computed tidal radii (see discussion in Section \ref{tirad}), and the
observed limiting radius, given in \citet{H96}.  Table \ref{tbl-2},
for the non-axisymmetric potential, is similar; but since neither the
orbital energy, $E$, nor the z- component of angular momentum, $h$,
(both computed in the inertial galactic frame) are constants of
motion, we give in the ninth and tenth columns the minimum and maximum
values attained by $h$ in the course of the complete orbit.  In order
to assess the influence of observational uncertainties on our results,
we have listed for each system two additional lines, corresponding to
the orbits of maximum and minimum energy as discussed at the end of
this section.

Many orbits are not noticeably affected by the bar. This is true for
all 'outer' clusters, i.e., those with orbits residing entirely
outside the bar region. The two left columns of Figure \ref{fig1} show
the meridional orbits of NGC 1904, NCG 2298, NGC 4590 and Pal 12 as
examples of this situation.  A similar behavior is exhibited by the
orbits of NGC 104, NGC 1851, NGC 5024, NGC 5272, NGC 5466, Pal 5, NGC
5904, NGC 6205, NGC 6934, NGC 7006, NGC 7089 and Pal 13.

A completely different situation occurs for orbits residing within, or
almost within, the bar region, as was already noted by
\citet{PMM04}. The orbits show great changes, tending to reach higher
values of apogalactic distance and distance to the galactic plane.
The two right columns of Figure \ref{fig1} and the two left columns of
Figure \ref{fig2} show as examples the orbits of NGC 5139, NGC 6171,
NGC 6266, NGC 6362, NGC 6254, NGC 6316, NGC 6528, and NGC 6723.  A
similar behavior is shown by the orbits of NGC 6093, NGC 6121, NGC
6144, NGC 6218, NGC 6304, NGC 6522, NGC 6626, NGC 6712, and NGC 6809.
The orbits of NGC 6522, NGC 6752, and NGC 6838 (these last two having
their perigalactic points near the boundary of the bar region) reach
considerably larger values of apogalactic distance in the presence of
the bar than in the axisymmetric case.  NGC 6304, NGC 6316, and NGC
6723 have near-resonant orbits in the axisymmetric potential, but get
pushed towards greater apogalactic distance by the bar.

Other interesting effects of the bar on the orbits are illustrated in
the two right columns of Figure \ref{fig2}. We show the orbits of NGC
7078, NGC 7099, NGC 288, and NGC 6218. In the case of NGC 7078 and NGC
7099, the bar tends to trap the orbit into a near-resonance. The
opposite effect can also occur, as seen in the case of NGC 288. Here,
the bar pushes the orbit away from a near-resonance. Something similar
happens for the orbits of NGC 6553 and NGC 6779. Also shown in Figure
\ref{fig2} is the case of NGC 6218, an orbit in near-resonance in the
axisymmetric potential, and fully disordered in the presence of the
bar.

Finally, the orbit of Pal 3 may be an escape orbit when run with the
available data.  An improvement of the observed parameters -distance
and proper motion- will most probably yield a normal, bound, orbit.

\begin{figure}
\plotone{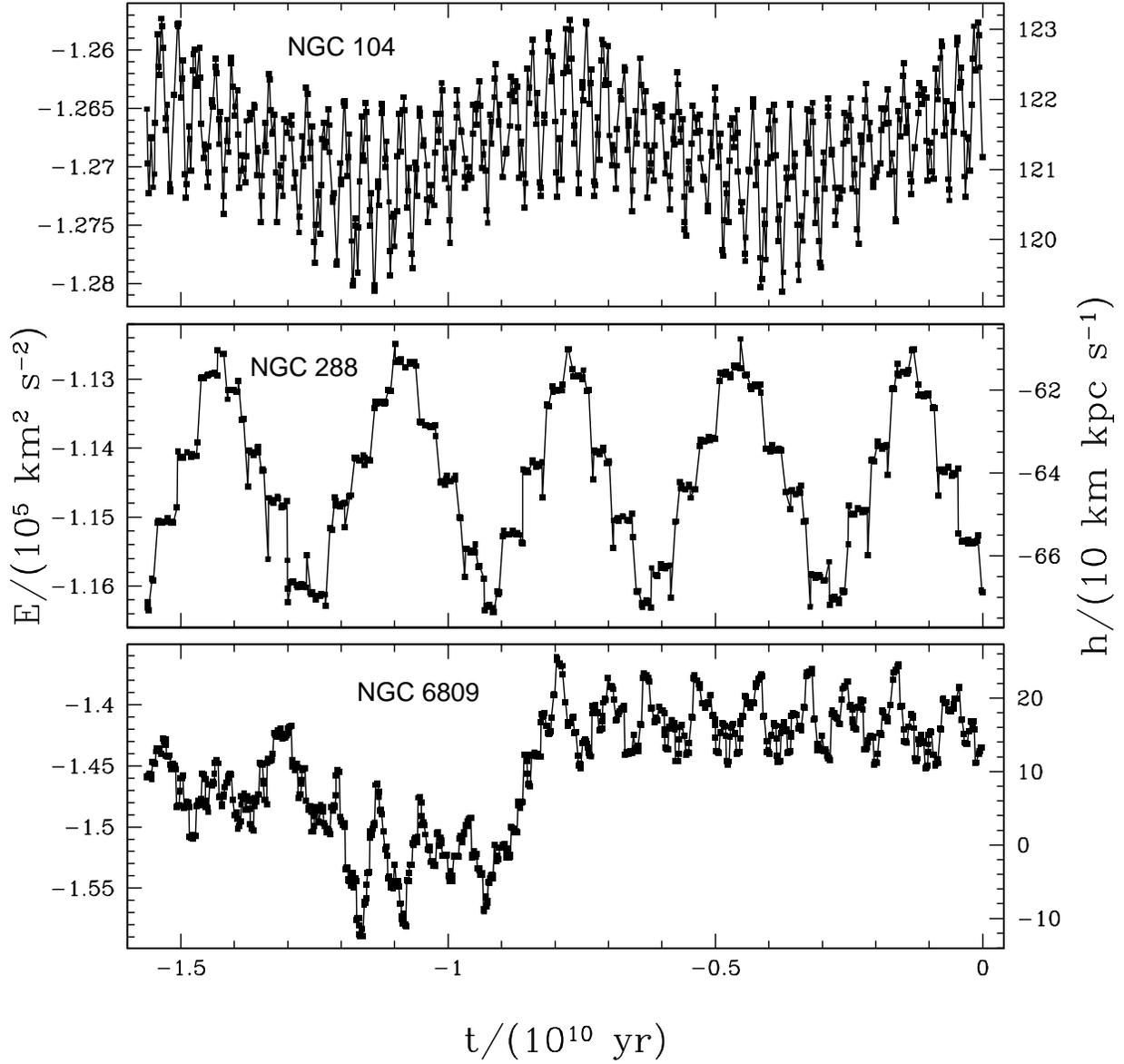}
\caption {Energy and z-angular momentum as functions of time in NGC
104, NGC 288, and NGC 6809, with the non-axisymmetric potential. $E$
and $h$ are connected by Jacobi's constant $J$: $J = E - {\Omega}h$;
$\Omega$ is the angular velocity of the bar and $h$ is positive in the
sense of galactic rotation.}
\label{fig3}
\end{figure}

Plots of the run of energy, $E$, and z-component of angular momentum,
$h$, were obtained for all clusters. These quantities are, of course,
conserved in the axisymmetric case. In the presence of the bar, the
orbits generally do not show secular changes in $E$ or $h$.  Indeed,
with a few exceptions, these quantities are conserved on the average
within better than 10 percent. Figure \ref{fig3} illustrates this
behavior for the cases of NGC 104 and NGC 288. However, in a few
cases, abrupt changes in $E$, $h$ do occur, as shown in Figure
\ref{fig3} for NGC 6809.  Here, on top of the nearly-periodic changes,
a jump occurs, moving the energy and angular momentum to higher values
(with increasing time) at the expense of the bar. Note that in the
case of NGC 6809, with an extremely low angular momentum orbit, the
abrupt changes in the angular momentum actually reverse the sense of
revolution of the orbit around the galactic center. Other examples of
this behavior are the orbits of NGC 6093, NGC 6316, and NGC 6528.

To take into account the influence observational uncertainties we
computed for each cluster two additional orbits in both the
axisymmetric potential and the time-varying barred potential.  Initial
conditions for these orbits were chosen so as to maximize and minimize
the energy.  In other words, we combined the observational
uncertainties in such a way as to obtain two extreme orbits. Errors in
the computed orbital parameters resulting from observational
uncertainties are expected to be bounded by these extreme orbits.  The
real uncertainties in our computed results will be much smaller than
those shown by the extreme orbits.  In Tables 1 and 2 we display two
additional lines for each system, corresponding to the minimum and the
maximum energy orbits, in both the axisymmetric and the barred
potential.  As can be seen from a comparison of the three orbits, the
errors in the computed parameters are quite moderate, except for NGC
1904, NGC 2298, Pal 3, NGC 4147, and NGC 7006.  The parameters most
affected by errors are $(r_{min})_{min}$, $<r_{min}>$, E and
h. Interestingly, Pal 3, the only escape orbit we found, becomes bound
in the minimum energy case, whereas NGC 7006 becomes unbound in the
maximum energy case.  So, within the observational uncertainties, all
our computed orbits are bound, as would be expected.  Of course, it
would be particularly valuable to obtain improved proper motions and
distances for the clusters showing the largest uncertainties.

In  Section 4 below we discuss the influence of observational 
uncertainties  on the computed tidal radii.

\section{TIDAL RADII}\label{tirad}

As a first step in assessing the effects of the bar on the internal
dynamics of the clusters, we have computed tidal radii.  The correct
formula to use for the purpose of comparing the theoretical tidal
radii with the observed limiting radii is a matter of some controversy
\citep[and references therein]{BOG99}. We are well aware of the fact
that a `tidal radius' is only a very coarse approximation to the
dynamical processes occurring in a cluster.  In every cluster we have
computed the tidal radius given by King's (1962) formula
$r_K=[M_c/M_g(3+e)]^{1/3}r_{min}$, with $M_c$ the cluster's mass,
$M_g$ an 'effective galactic mass', $e$ the orbital eccentricity, and
$r_{min}$ the galactocentric distance of a perigalactic point in the
orbit. As discussed in Appendix A, we also compute the tidal radius of
a cluster using an alternative approximate formula. This formula is
given by Eq. (\ref{x'gal}),

\begin{equation} r_{\ast} = \left [ \frac{GM_c}{\left (\frac{\partial
F_{x'}}
{\partial x'} \right )_{{\bf r'}= 0}+ \dot{\theta}^2 + \dot{\varphi}^2\sin^2
{\theta}} \right ]^{1/3}. \label{rast} \end{equation}

In this equation, $F_{x'}$ is the component of the galactic
acceleration along the line $x'$ joining the cluster with the galactic
center, and its partial derivative is evaluated at the position of the
cluster. The angles ${\varphi}$ and ${\theta}$ are angular spherical
coordinates of the cluster in an inertial galactic frame. The tidal
radius $r_{\ast}$ is computed at a perigalactic point in the orbit. In
a two-body problem, $r_{\ast}$ coincides with $r_K$ given by King's
(1962) formula. The main advantage in the computation of $r_{\ast}$ is
that we do not need to make an assumption about an 'effective galactic
mass'; an assumption which is unclear if the perigalactic point lies
in a region with a strong non-axisymmetric potential, as in the case
of a bar.

\begin{figure}
\plotone{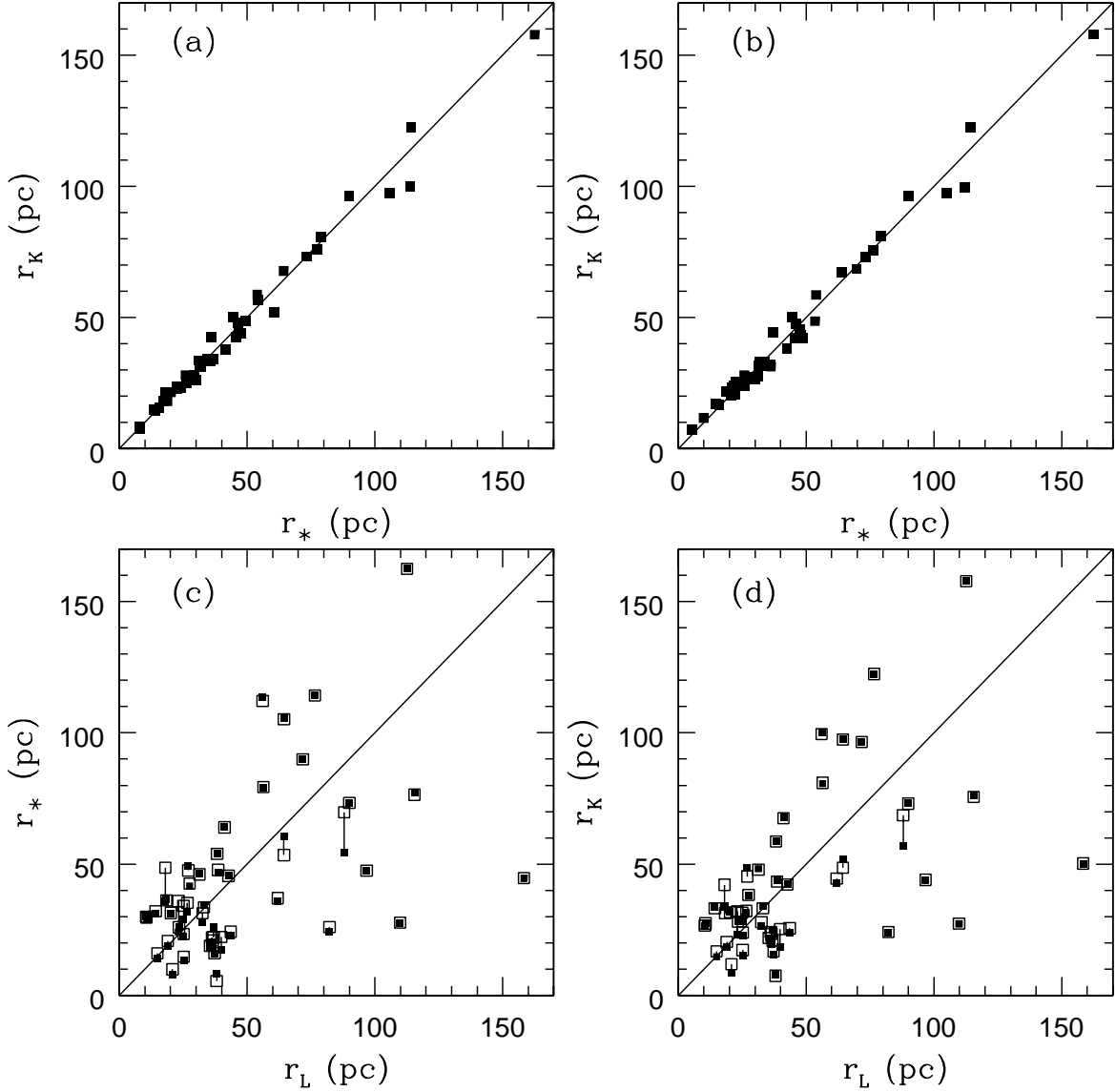}
\caption {Comparison of tidal radii.  The tidal radii $r_{K}$ and
$r_{\ast}$ are compared using (a) the axisymmetric, and (b) the
non-axisymmetric galactic potentials. In (c) and (d), $r_{\ast}$ and
$r_{K}$ are compared with the observed limiting radius $r_{L}$; filled
and empty squares are obtained with the axisymmetric and
non-axisymmetric potentials, respectively. In each frame we plot the
line of coincidence.}
\label{fig4}
\end{figure}

\begin{figure}
\plotone{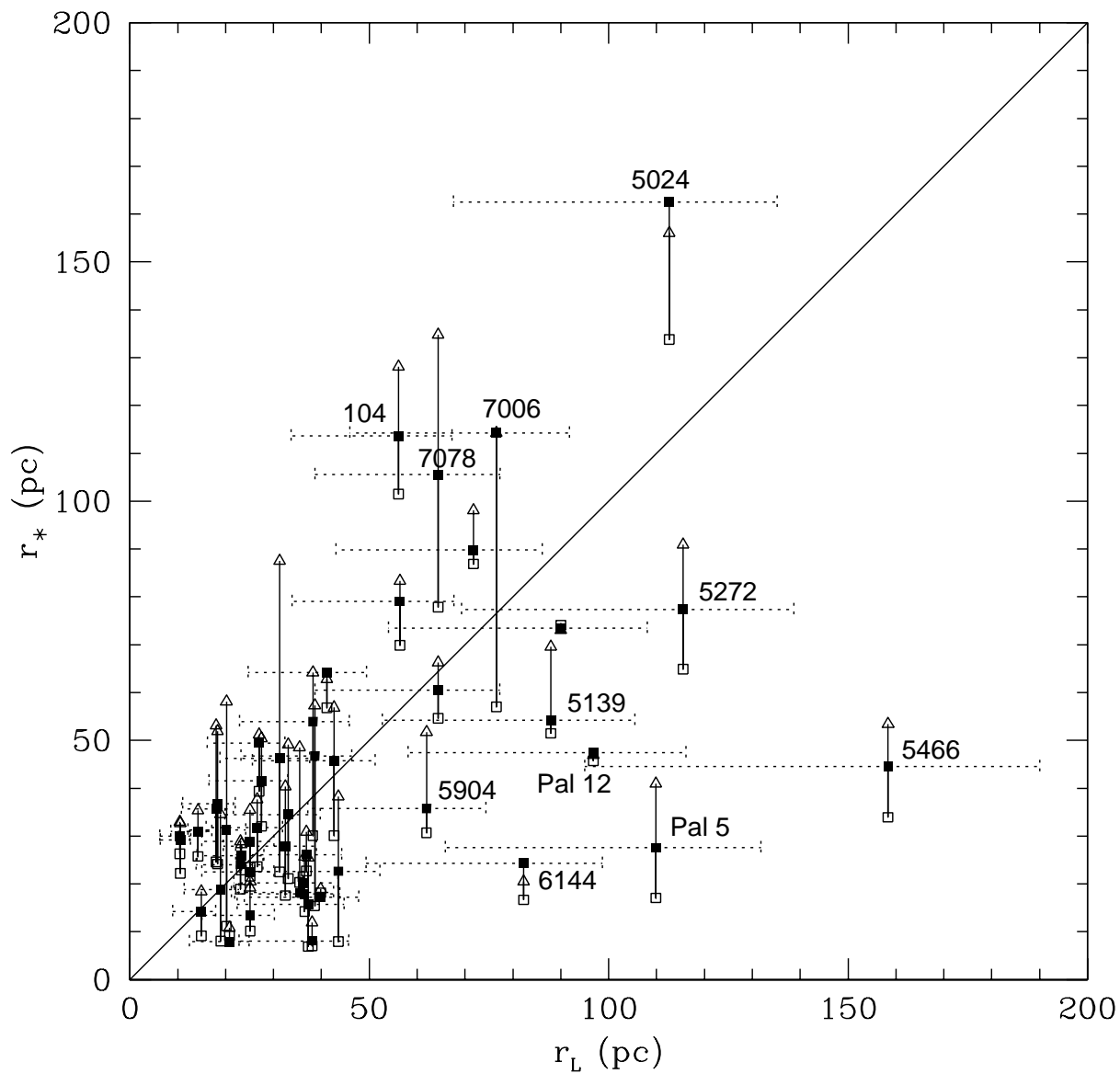}
\caption {Comparison of the tidal radii $r_{\ast}$ and $r_{L}$ using the
axisymmetric potential. For each cluster we plot three points (filled squares,
empty squares and empty triangles) corresponding respectively to three orbits:
the 'central' orbit and the two extreme orbits of minimum and maximum energy.
They appear joined by vertical lines, as 'error bars'. The horizontal dotted
lines give the estimated observational uncertainties in $r_{L}$. Some NGC and 
Pal numbers are shown. The line of coincidence is plotted.}
\label{fig5}
\end{figure}

\begin{figure}
\plotone{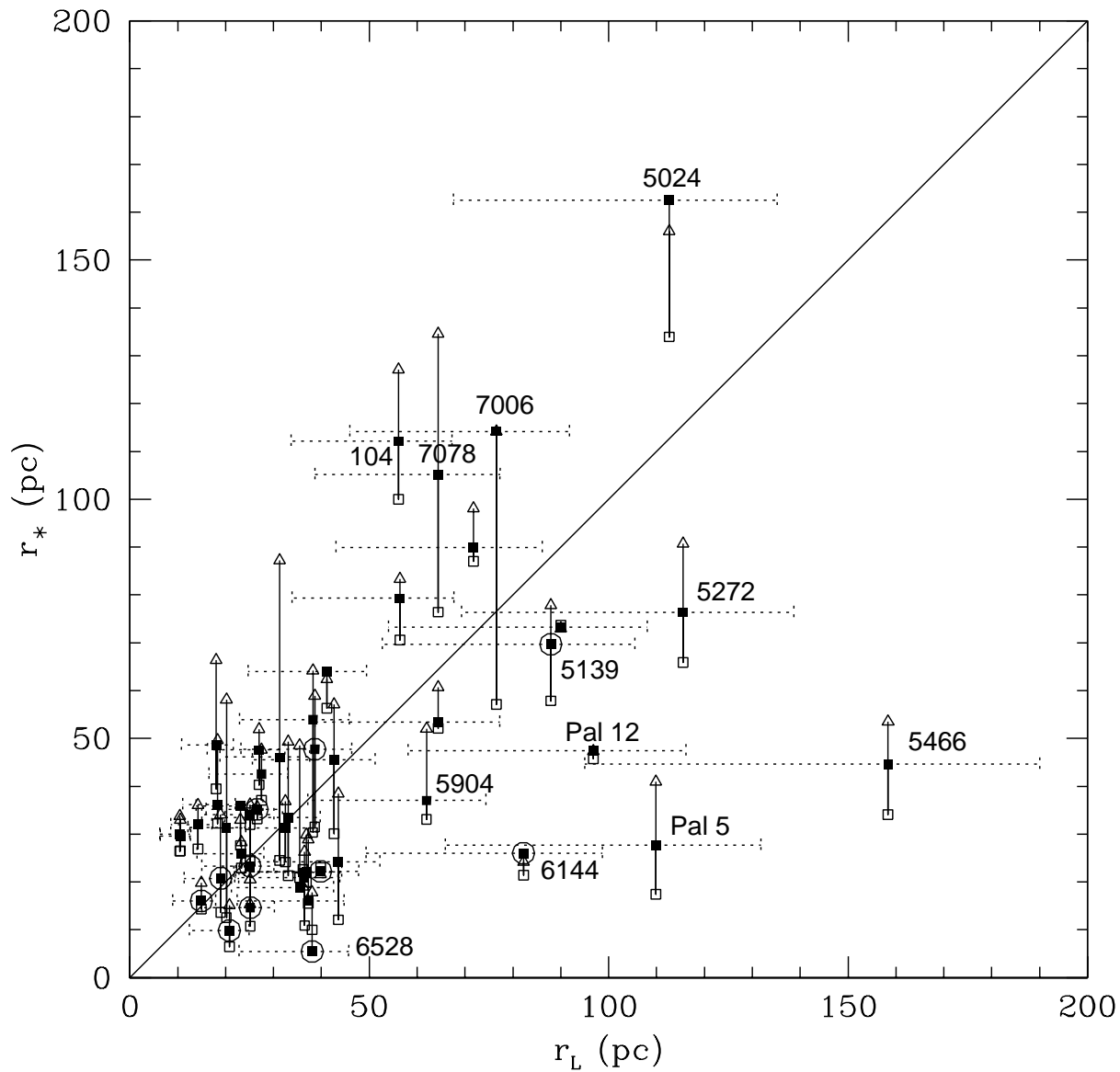}
\caption {As in Figure \ref{fig5}, but now using the non-axisymmetric
potential. The central, minimum, and maximum energy orbits are
computed with the initial (t = 0) conditions of the corresponding
orbits in the axisymmetric potential. The filled squares with a circle
correspond to clusters with a retrograde galactic orbit, or with an
orbit which is both prograde and retrograde, and with a mean
perigalactic distance less than 3 kpc.}
\label{fig6}
\end{figure}

Figure \ref{fig4} shows the comparison of the tidal radii $r_{K}$ and
$r_{\ast}$, and their relation with the observed limiting radius
$r_{L}$.  In Figure \ref{fig4} (a), (b) we compare $r_{K}$ and
$r_{\ast}$ in the axisymmetric and non-axisymmetric potentials,
respectively. Each point corresponds to a cluster, and we are plotting
the averages of the tidal radii over the last $10^9$ yr in every
cluster's orbit. These frames show that $r_{K}$ and $r_{\ast}$ give
similar values. This conclusion is maintained if we compare $r_{K}$
and $r_{\ast}$ in a given orbit at any perigalactic point.  In Figure
\ref{fig4} (c), (d) we compare $r_{\ast}$ and $r_{K}$ with the
observed limiting radius $r_{L}$. Both frames include the results
obtained with the axisymmetric (filled squares) and non-axisymmetric
(empty squares) potentials. In general, the values of $r_{\ast}$,
$r_{K}$ obtained with the non-axisymmetric potential are almost the
same as those obtained with the axisymmetric potential (this can also
be seen comparing (a) and (b) of Figure \ref{fig4}). We also note that
both frames are very similar (because $r_{\ast}$ and $r_{K}$ are
similar), and there is a considerable scatter, mainly in the lower
right-hand side in these frames. In order to discuss this scatter, in
Figure \ref{fig5} we compare $r_{\ast}$ with $r_{L}$, but showing only
the results obtained in the axisymmetric potential. To estimate the
lower and upper bounds in $r_{\ast}$, we have computed for each
cluster the minimum and maximum energy orbits permitted by the
available data. These orbits give the empty squares and empty
triangles in Figure \ref{fig5}. We also show in this figure the
estimated observational uncertainties in $r_{L}$. Three outliers
contribute most of the noise: the points lying in the lower right-hand
side of the diagram. As shown in the figure, these points correspond
to NGC 6144, Pal 5, and NGC 5466.  All three are low-mass,
low-concentration clusters. Pal 5 appears to be in the final stages of
dissolution \citep{KGOMC04}, and there are indications that NGC 5466
is also dissolving \citep{OG04,B06}.  The case of NGC 6144 is
commented below, in relation with Figure \ref{fig6} for the
non-axisymmetric potential.  Thus, in the cases of Pal 5 and NGC 5466
the present limiting radii of these clusters do not necessarily
correspond to the conditions of the dynamical model for the
theoretical tidal radii (which assumes the cluster has undergone a
recent episode of tidal pruning). Other clusters lying significantly
below the line of coincidence in Figure \ref{fig5}, are NGC 5272, NGC
5139, and NGC 5904, for all three of which tidal tails have been
detected \citep{LMC00}, thus indicating a similar situation to that of
Pal 5, if not as extreme. Therefore, agreement between the theoretical
tidal radii and the observed limiting radii is not to be expected for
these clusters either, as indeed is not observed.

Interestingly, most of the clusters lying significantly above the line
of coincidence (NGC 1851, NGC 1904, NGC 2298, NGC 5024, NGC 6205, and
NGC 7089) have been proposed as possibly being accreted systems,
because they are old clusters with young halo kinematics
\citep{MG04}. This is also the case of NGC 5904 and Pal 12, clusters
that lie below the line of coincidence; \citet{D00} find that Pal 12
has probably been captured from the Sagittarius dwarf galaxy.  A few
of the clusters that do not fill their tidal radii (NGC 1851, NGC
1904, and NGC 6205) also exhibit tidal tails \citep{LMC00}. Taking
into account the preceeding considerations, we can say that in general
the agreement between the computed and the observed tidal radii of
'normal' clusters is quite good. The outstanding exception is NGC 104,
whose observed limiting radius is much smaller than its tidal radius,
i.e., for no reason that we can discern, this cluster does not fill
its tidal radius.

Figure \ref{fig6} shows the comparison of $r_{\ast}$ with $r_{L}$ in
the non-axisymmetric potential. The details in this figure are
analogous to those in Figure \ref{fig5}; additionally, the points with
a circle correspond to clusters with a retrograde galactic orbit, or
with an orbit which is both prograde and retrograde, and with a mean
perigalactic distance less than 3 kpc. Only in the galactic
non-axisymmetric potential, which has a rotating bar, the sense of
orbital rotation around the galactic center acquires an
importance. Some of the clusters with a retrograde galactic orbit and
mean perigalactic distance within the bar region, and which do not lie
near the line of coincidence in Figure \ref{fig6}, may partly do so
because in the computation of a tidal radius in this non-axisymmetric
potential, in principle the sense of orbital rotation must be
included, which is not the case in $r_{K}$ and
$r_{\ast}$. \citet{KI75} showed that stars within a cluster moving in
retrograde orbits are more stable in a tidal field than stars moving
in prograde orbits. This effect is not included in the computation of
$r_{K}$ and $r_{\ast}$ either. Figure \ref{fig6} suggests that
clusters moving in retrograde orbits in the galactic non-axisymmetric
potential are more stable to the tidal field than clusters moving in
prograde orbits; this is because several points corresponding to
retrograde orbits lie below the line of coincidence, that is, $r_{L}$
is greater than $r_{\ast}$ (or $r_{K}$).  Examples of this effect are
the orbits of NGC 6144 and NGC 6528 (see Figure \ref{fig6}); note that
NGC 6528 is the cluster with the smallest computed theoretical tidal
radius.

\section{DESTRUCTION RATES}\label{destr}

We summarize in Sections \ref{bulbo} and \ref{disco} some equations
employed to compute destruction rates of globular clusters due to
gravitational shocks produced by a galactic bulge and a galactic
disk. In Section \ref{destrres} we apply this theory to globular
clusters in our Galaxy.

Destruction rates of globular clusters in an axisymmetric model of our
Galaxy were obtained in the pioneer work of \citet{AHO88}, who
considered evaporation due to two-body relaxation, dynamical friction,
and gravitational shocks with both the galactic bulge and galactic
disk. The effect of a galactic bar on these destruction rates was
analyzed by \citet{LOA92}. For gravitational shocks, \citet[hereafter
G\&O97]{GO97} considered the effect of the energy change $<$$\Delta
E$$>$ as well as of the second-order term $<$$(\Delta E)^2$$>$, on a
cluster's evolution, introducing in each case adiabatic corrections to
the impulse approximation with a power-law form \citep[hereafter
G\&O99]{GO99}, as improvements to the exponential forms given by
\citet{S87} and \citet{KO95}. \citet{G99a} also improved the treatment
computing the tidal field due to an extended spherically symmetric
mass distribution, as in a spherical galactic bulge.

\subsection{Bulge Shock}\label{bulbo}

When a globular cluster reaches a perigalactic point with position
{\boldmath $r$}$_p$ and velocity {\boldmath $v$}$_p$ with respect to
the galactic inertial frame, the gravitational shock with the bulge,
of total mass $M_b$, produces a change in binding energy per unit mass
given by (taking an average over the cluster of equation (17) in
G\&O97)

\begin{equation} <\!(\Delta E)_b\!> = \frac{1}{3} \left (\frac{2GM_b}
{|\mbox{\boldmath $v$}_p||\mbox{\boldmath $r$}_p|^2} \right )^2
{\chi}(|\mbox{\boldmath $r$}_p|)<\!r'^2 {\eta}_1 (\beta(r'))\!> \lambda
(|\mbox{\boldmath $r$}_p|, |\mbox{\boldmath $r$}_a|).
\label{delEb} \end{equation}

The factor ${\chi}(|${\boldmath $r$}$_p|)$ takes into account the
extended mass distribution of the bulge, and is given in equation (18)
of G\&O97 (see also equation (17) of \citet{G99a}).  In Appendix B we
show that the expression they give for this factor can be simplified,
and needs the computation of only two integrals over the mass
distribution of the bulge. Thus, using the results obtained in
Appendix B we have

\begin{equation} {\chi}(|\mbox{\boldmath $r$}_p|) = \frac{1}{2} \left [
{\Im}_1^2(|\mbox{\boldmath $r$}_p|) + \left ( 1 - \frac{4\pi|
\mbox{\boldmath $r$}_p|^3}{M_b}{\Im}_2(|\mbox{\boldmath $r$}_p|)
\right )^2 \right ], \label{fchi} \end{equation}

\noindent with

\begin{equation} {\Im}_1(|\mbox{\boldmath $r$}_p|) = \int_{1}^{\infty}
\frac{{\mu}_b(u|\mbox{\boldmath $r$}_p|)du}{u^2(u^2-1)^{1/2}},
\label{int1} \end{equation}

\begin{equation} {\Im}_2(|\mbox{\boldmath $r$}_p|) = \int_{1}^{\infty}
\frac{{\rho}_b(u|\mbox{\boldmath $r$}_p|)u^3du}{(u^2-1)^{1/2}}.
\label{int2} \end{equation}

\noindent Here ${\mu}_b(r) = M_b(r)/M_b$, with $M_b(r)$ the mass of
the bulge within a sphere of galactocentric radius $r$, $M_b$ is the
total mass of the bulge, and ${\rho}_b(r)$ is the density of the
spherically symmetric bulge as a function of galactocentric distance
$r$.

The factor $<$$r'^2 {\eta}_1 (\beta(r'))$$>$ in Eq. (\ref{delEb}) is
an average over the cluster; $r'$ is distance from the center of the
cluster, $\beta (r')$ is the adiabatic parameter $\beta (r') =
2|$\mbox{\boldmath $r$}$_p| \omega (r')/|$\mbox{\boldmath $v$}$_p|$,
with $\omega (r')$ angular velocity of stars inside the
cluster. Considering circular motions as representative, then $\omega
(r') = [GM_c(r')/r'^3]^{1/2}$, with $M_c(r')$ the mass of the cluster
within a sphere of radius $r'$. The function ${\eta}_1 (\beta)$
($A_{b1}$ in the notation of G\&O97) is the adiabatic correction of
the form given by G\&O99: ${\eta}_1 (\beta) = (1 +
\frac{1}{4}{\beta}^2)^{-{\gamma}_1}$ ($\beta = 2x$ in their
notation). The exponent ${\gamma}_1$ is given in Table 2 of G\&O99,
and we extended its values as follows: ${\gamma}_1$ = 2.5 if $\tau
\equiv |${\boldmath $r$}$_p|/|${\boldmath $v$}$_p|\leq t_{dyn,h}$,
${\gamma}_1$ = 2 if $t_{dyn,h} < \tau < 4t_{dyn,h}$, ${\gamma}_1$ =
1.5 if $\tau \geq 4t_{dyn,h}$, with $t_{dyn,h} = ({\pi}^2
{r'_h}^2/2GM_c)^{1/2}$ the half-mass dynamical time, where $r'_h$ is
the half-mass radius and $M_c$ the mass of the cluster. The average
$<$$r'^2 {\eta}_1 (\beta(r'))$$>$ is taken as

\begin{equation} <\!r'^2 {\eta}_1 (\beta(r'))\!> = \frac{4\pi}{M_c}
\int_{0}^{r'_t} {\rho}_c(r'') {\eta}_1(\beta(r'')) r''^4 dr'', \label{prom1}
\end{equation}

\noindent with $r'_t$ the tidal radius of the cluster and ${\rho}_c(r')$ its
spatial density, obtained with a \citet{K66} model.

The factor $\lambda (|${\boldmath $r$}$_p|, |${\boldmath $r$}$_a|)$ in
Eq. (\ref{delEb}), a function of the perigalactic, $|${\boldmath
$r$}$_p|$, and apogalactic, $|${\boldmath $r$}$_a|$, distances, is an
approximation for the effect of the time variation of the galactic
tidal force along the orbit of the cluster. According to \citet{AHO88}
this factor can be taken as

\begin{equation} \lambda (|\mbox{\boldmath $r$}_p|, |\mbox{\boldmath $r$}_a|)
= \left [1- \frac{M_b(|\mbox{\boldmath $r$}_a|)}{M_b(|\mbox{\boldmath $r$}_p|)}
\left (\frac{|\mbox{\boldmath $r$}_p|}{|\mbox{\boldmath $r$}_a|} \right )^3
\right ]^2. \label{lambda} \end{equation}

The second-order term $<$$(\Delta E)_b^2$$>$ in the gravitational
shock with the bulge is given by equation (19) of \citet{G99a},
modified by the factors ${\chi}(|${\boldmath $r$}$_p|)$, $\lambda
(|${\boldmath $r$}$_p|, |${\boldmath $r$}$_a|)$ and the appropriate
adiabatic correction factor ${\eta}_2 (\beta(r'))$. Thus, an average
over the cluster gives

\begin{equation} <\!(\Delta E)_b^2\!> = \frac{2}{9} \left (\frac{2GM_b}
{|\mbox{\boldmath $v$}_p||\mbox{\boldmath $r$}_p|^2} \right )^2
{\chi}(|\mbox{\boldmath $r$}_p|)<\!r'^2 v'^2 {\eta}_2 (\beta(r'))
(1 + {\chi}_{r',v'})\!> \lambda(|\mbox{\boldmath $r$}_p|,
|\mbox{\boldmath $r$}_a|),
\label{delE2b} \end{equation}

\noindent and we have approximated the rms velocity $v'$ within the
cluster with that given by circular motion; thus $r'^2v'^2 =
{\omega}^2(r')r'^4 = GM_c(r')r'$ (see equation (22) of G\&O97). The
adiabatic correction is ${\eta}_2 (\beta) = (1 +
\frac{1}{4}{\beta}^2)^{-{\gamma}_2}$, with ${\gamma}_2$ given in Table
2 of G\&O99 and extended as follows: ${\gamma}_2$ = 3 if $\tau \equiv
|${\boldmath $r$}$_p|/|${\boldmath $v$}$_p|\leq t_{dyn,h}$,
${\gamma}_2$ = 2.25 if $t_{dyn,h} < \tau < 4t_{dyn,h}$, ${\gamma}_2$ =
1.75 if $\tau \geq 4t_{dyn,h}$. The position-velocity correlation
function ${\chi}_{r',v'}$ is given by G\&O99, and the average $<$$r'^2
v'^2 {\eta}_2 (\beta(r'))(1 + {\chi}_{r',v'})$$>$ is

\begin{equation} <\!r'^2v'^2 {\eta}_2 (\beta(r'))(1 + {\chi}_{r',v'})\!> =
\frac{4{\pi}G}{M_c}
\int_{0}^{r'_t} {\rho}_c(r'') M_c(r'') {\eta}_2(\beta(r''))
(1+{\chi}_{r',v'}(r'')) r''^3 dr''. \label{prom2} \end{equation}

Following G\&O97, bulge shock timescales are defined as

\begin{equation} t_{bulge,1} = \left ( \frac{-E_c}{<\!(\Delta E)_b\!>}
\right )P_{orb}, \label{tb1} \end{equation}

\begin{equation} t_{bulge,2} = \left ( \frac{E_c^2}{<\!(\Delta E)_b^2\!>}
\right )P_{orb}, \label{tb2} \end{equation}

\noindent with $E_c \simeq -0.2GM_c/r'_h$ the mean binding energy per
unit mass of the cluster and $P_{orb}$ its orbital period in the
galactic radial direction. The total destruction rate due to
gravitational shocks with the bulge is

\begin{equation} \frac{1}{t_{bulge}} = \frac{1}{t_{bulge,1}} +
\frac{1}{t_{bulge,2}}. \label{tbtot} \end{equation}

\subsection{Disk Shock}\label{disco}

For a gravitational shock with a galactic disk, the corresponding
expressions of $<$$(\Delta E)_d$$>$ and $<$$(\Delta E)_d^2$$>$ in a
cluster are obtained with averages of equations (1) and (2) in
\citet{G99b}:

\begin{equation} <\!(\Delta E)_d\!> = \frac{2g_m^2}{3v_z^2}
<\!r'^2 {\eta}_1 (\beta(r'))\!>, \label{delEd} \end{equation}

\begin{equation} <\!(\Delta E)_d^2\!> = \frac{4g_m^2}{9v_z^2}
<\!r'^2 v'^2 {\eta}_2 (\beta(r'))(1 + {\chi}_{r',v'})\!>,
\label{delE2d} \end{equation}

\noindent with $|g_m|$ the maximum acceleration produced by the disk
in the z-direction, perpendicular to its plane, on the perpendicular
line at the position where the galactic orbit of the cluster passes
through the disk; $|v_z|$ is the z-velocity of the cluster at this
point. The adiabatic functions ${\eta}_1(\beta)$, ${\eta}_2(\beta)$
are those employed in the gravitational shock with a bulge, but now
$\beta(r') = 2|z_m|\omega(r')/|v_z|$, with $|z_m|$ the height above
the plane of the disk where $|g_m|$ is reached. As before,
$\omega(r')$ is angular velocity of circular motions within the
cluster. Also, the parameter $\tau$, whose value gives the
corresponding values of the exponents ${\gamma}_1, {\gamma}_2$ in the
adiabatic corrections, is $\tau = |z_m|/|v_z|$. The averages in
Eqs. (\ref{delEd}), (\ref{delE2d}) are obtained with Eqs.
(\ref{prom1}), (\ref{prom2}).

If the orbit of the cluster has $n$ crossings (not necessarily two)
with the disk during the radial orbital period $P_{orb}$, disk shock
timescales and total destruction rate due to gravitational shocks with
the disk are defined as

\begin{equation} t_{disk,1} = \left ( \frac{-E_c}{<\!(\Delta E)_d\!>}
\right ) \frac{P_{orb}}{n}, \label{td1} \end{equation}

\begin{equation} t_{disk,2} = \left ( \frac{E_c^2}{<\!(\Delta E)_d^2\!>}
\right ) \frac{P_{orb}}{n}, \label{td2} \end{equation}

\begin{equation} \frac{1}{t_{disk}} = \frac{1}{t_{disk,1}} +
\frac{1}{t_{disk,2}}. \label{tdtot} \end{equation}

\subsection{Destruction rates of globular clusters in our Galaxy}\label{destrres}

\begin{figure}
\plotone{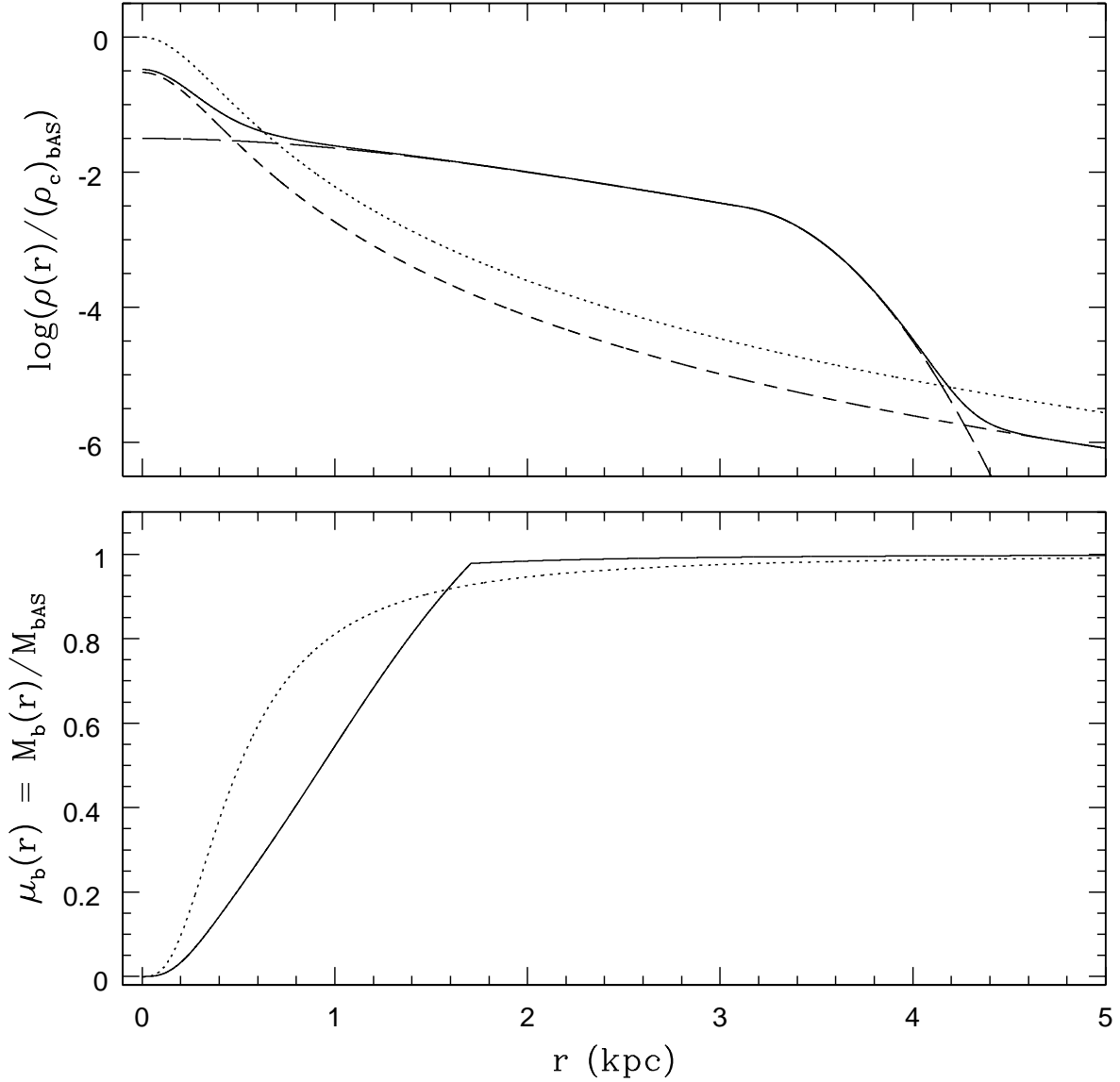}
\caption {The 'bulge' in the non-axisymmetric galactic potential. The
upper frame shows the density of the bar and the
spherical bulge, scaled by the central density of the bulge in the
axisymmetric potential, as functions of galactocentric distance. The
dotted line shows the density of the bulge in the axisymmetric
galactic potential; the short-dashed line shows the density of the
spherical bulge, and the long-dashed line the density of
the galactic bar along its major axis. The continuous line gives the
total density of both components. For the approximate computation in
the non-axisymmetric potential of ${\chi}(|${\boldmath $r$}$_p|)$ and
$\lambda (|${\boldmath $r$}$_p|, |${\boldmath $r$}$_a|)$ in
Eqs. (\ref{fchi}) and (\ref{lambda}), in the lower frame the
continuous line gives the mass within a sphere of galactocentric
radius $r$, scaled by the total mass of the bulge in the axisymmetric
galactic potential, of the 'equivalent' total bulge (see the
text). The dotted line in this frame corresponds to the bulge in the
axisymmetric potential.}
\label{fig7}
\end{figure}

The formulism given in the two previous sections has been applied to
globular clusters in our Galaxy, considering the axisymmetric and
non-axisymmetric galactic potentials. In the non-axisymmetric galactic
potential \citep{PMM04} the original spherical bulge in the
axisymmetric potential is left with a small fraction (30 \%) of its
mass, the rest (70 \%) goes into the galactic bar.  In the Cartesian
axes defined by the principal axes of the bar, the density of the bar
at a point ($x,y,z$) is approximately the density given by Model S of
\citet{F98}: ${\rho}_{Bar}(x,y,z) \propto sech^{2}(R_S(x,y,z))$ (see
\citet{PMM04}). In the upper frame of Figure \ref{fig7} we show the
density of the bar and the spherical bulge, as functions of
galactocentric distance and scaled by the central density of the bulge
in the axisymmetric galactic potential. The dotted line gives the
density of this bulge in the axisymmetric galactic potential; the
short-dashed line shows the density of the spherical bulge and the
long-dashed line gives the density of the galactic bar along its major
axis. The continuous line gives the total density of these two
components. Notice that the bar now gives a moderate steepness in the
run of density, as compared with the original bulge. In our
computations in the non-axisymmetric potential, we take as the 'bulge'
the galactic bar plus the spherical bulge; hence this bulge is not
spherically symmetric.  The computation of destruction rates due to
bulge and disk shocking is easily done in the axisymmetric galactic
potential. However, in the non-axisymmetric galactic potential the
computation of ${\chi}(|${\boldmath $r$}$_p|)$ and $\lambda
(|${\boldmath $r$}$_p|, |${\boldmath $r$}$_a|)$ in Eqs. (\ref{fchi})
and (\ref{lambda}) is not immediate, as in these factors the bulge is
supposed to be spherically symmetric. Thus, an approximation must be
done to compute these factors. \citet{LOA92}, who used a bar to
evaluate its effect on the destruction rates of globular clusters,
must also use some approximation to compute their factor ${\chi}$.

Our approximation to compute ${\chi}(|${\boldmath $r$}$_p|)$ and
$\lambda (|${\boldmath $r$}$_p|, |${\boldmath $r$}$_a|)$ in the
non-axisymmetric galactic potential, involves representing the 'bulge'
as spherical and transforming the galactic bar to an 'equivalent'
spherical bulge. The density of the bar, in terms of Cartesian
coordinates along its principal axes, ${\rho}_{Bar}(x,y,z) \propto
sech^{2}(R_S(x,y,z))$, is associated to the galactocentric distance $r
= \frac{1}{3}(a_x + a_y + a_z)R_S(x,y,z)$, with $a_x, a_y, a_z$ the
scale lengths of the bar \citep{F98}. In the lower frame of Figure
\ref{fig7} we show as a continuous line the ratio ${\mu}_b(r) =
M_b(r)/M_{bAS}$, with $M_b(r)$ the mass of the 'equivalent' total
bulge (that is, the spherical bulge plus the 'equivalent' spherical
bulge associated with the galactic bar) within a sphere of
galactocentric radius $r$; $M_{bAS}$ is the total mass of the bulge in
the axisymmetric galactic potential. The dotted line in this frame
corresponds to the bulge in the \citet{AS91} galactic
potential. Notice that even with this contracted-bulge approximation,
the 'equivalent' total bulge is not the same as the bulge in the
axisymmetric potential. We stress the fact that the above
approximation is used only to compute ${\chi}(|${\boldmath $r$}$_p|)$
and $\lambda (|${\boldmath $r$}$_p|, |${\boldmath $r$}$_a|)$; but a
cluster's galactic orbit is strictly computed in the non-axisymmetric
galactic potential.  Then $|${\boldmath $r$}$_p|$, $|${\boldmath
$v$}$_p|$, and $|${\boldmath $r$}$_a|$ employed in Section \ref{bulbo}
are known for any orbital period. In any case, the factor $\lambda
(|${\boldmath $r$}$_p|, |${\boldmath $r$}$_a|)$ is not at all well
determined \citep{AHO88}. A better treatment to compute the effect of
the bar, without the above approximation, is under way.

In Table \ref{tbl-3} we present the destruction rates obtained in our
computations. The first column shows the name of the cluster. The
second column gives the mass of the cluster, computed with a
mass-to-light ratio of 2. Absolute visual magnitudes are taken from
\citet{H96}. The only exception is NGC 5139 (Omega Cen), whose mass is
taken from \citet{PA75}. The central concentration and the half-mass
radius are given in columns three and four, and are taken from
\citet{H96}. The remaining columns show the rounded destruction rates
due to the bulge and disk (see Eqs. (\ref{tbtot}) and
(\ref{tdtot}). These destruction rates are the averages over the last
$10^9$ yr in a cluster's orbit, taking the corresponding perigalactic
points and crossings with the galactic disk; this time interval is
extended in some cases (NGC 1851, NGC 4147, NGC 4590, NGC 5024, NGC
5466, NGC 5904, NGC 6205, NGC 6934, NGC 7006, NGC 7089, Pal 12, and
Pal 13) to allow for a representative number of perigalactic points
and crossings with the disk. For a given cluster, the first line in
Table \ref{tbl-3} shows the destruction rates computed in the
axisymmetric potential; the second line gives the corresponding values
in the non-axisymmetric potential.

\begin{figure}
\plotone{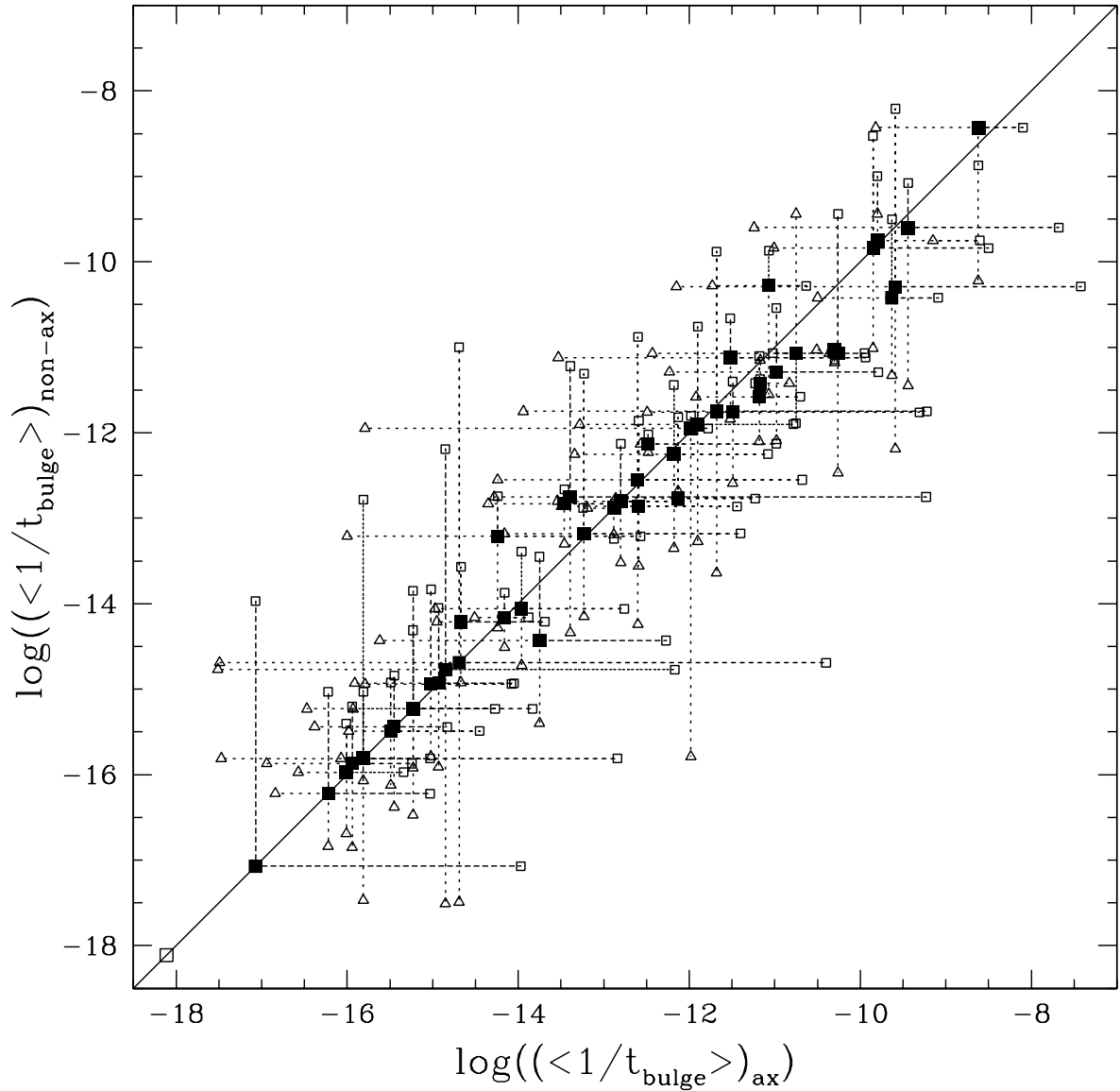}
\caption {Comparison of the destruction rates due to the bulge, in the
  axisymmetric and non-axisymmetric potentials. Values obtained with
  the 'central', minimum energy, and maximum energy orbits are shown
  as filled squares, empty squares, and empty triangles, respectively.
  The bottom left empty square corresponds to Pal 3, whose only bound
  orbit is that of minimum energy. The line of coincidence is
  plotted.}
\label{fig8}
\end{figure}

\begin{figure}
\plotone{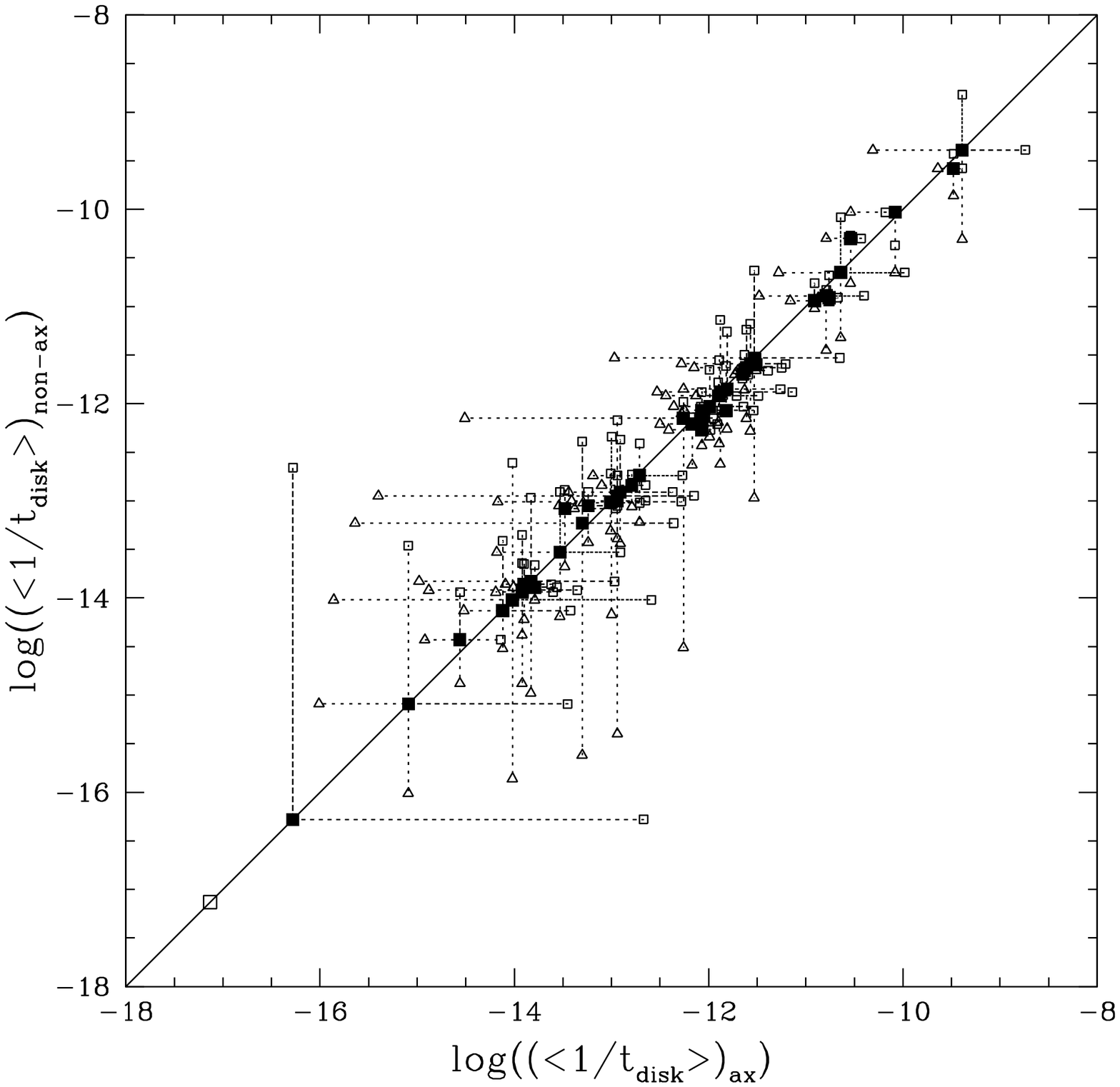}
\caption {Comparison of the destruction rates due to the disk, in the
axisymmetric and non-axisymmetric potentials. Points as in Figure \ref{fig8}.}
\label{fig9}
\end{figure}

\begin{figure}
\plotone{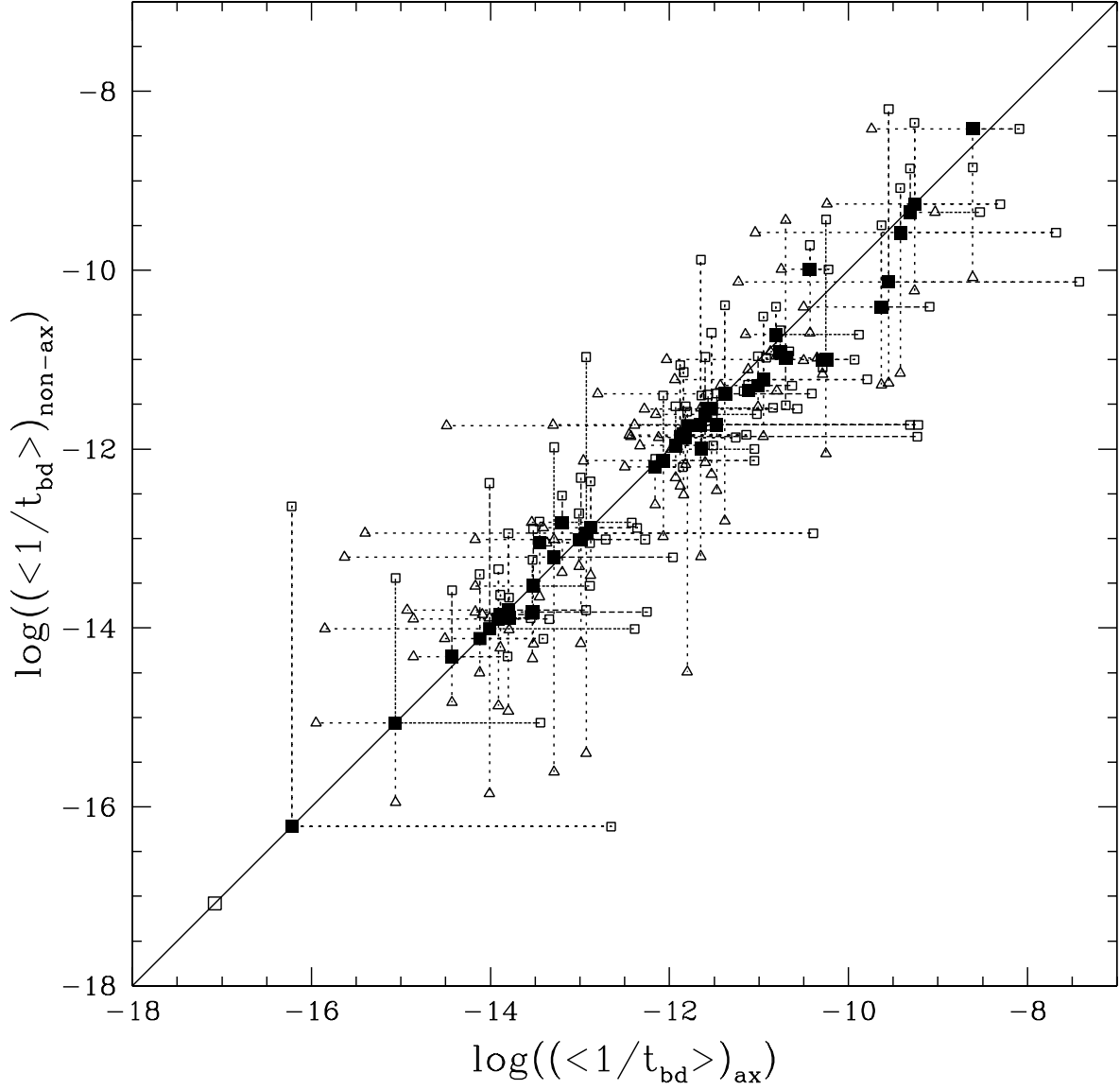}
\caption {Comparison of the total destruction rates due to the bulge and disk,
in the axisymmetric and non-axisymmetric potentials. 
Points as in Figure \ref{fig8}.}
\label{fig10}
\end{figure}

\begin{figure}
\plotone{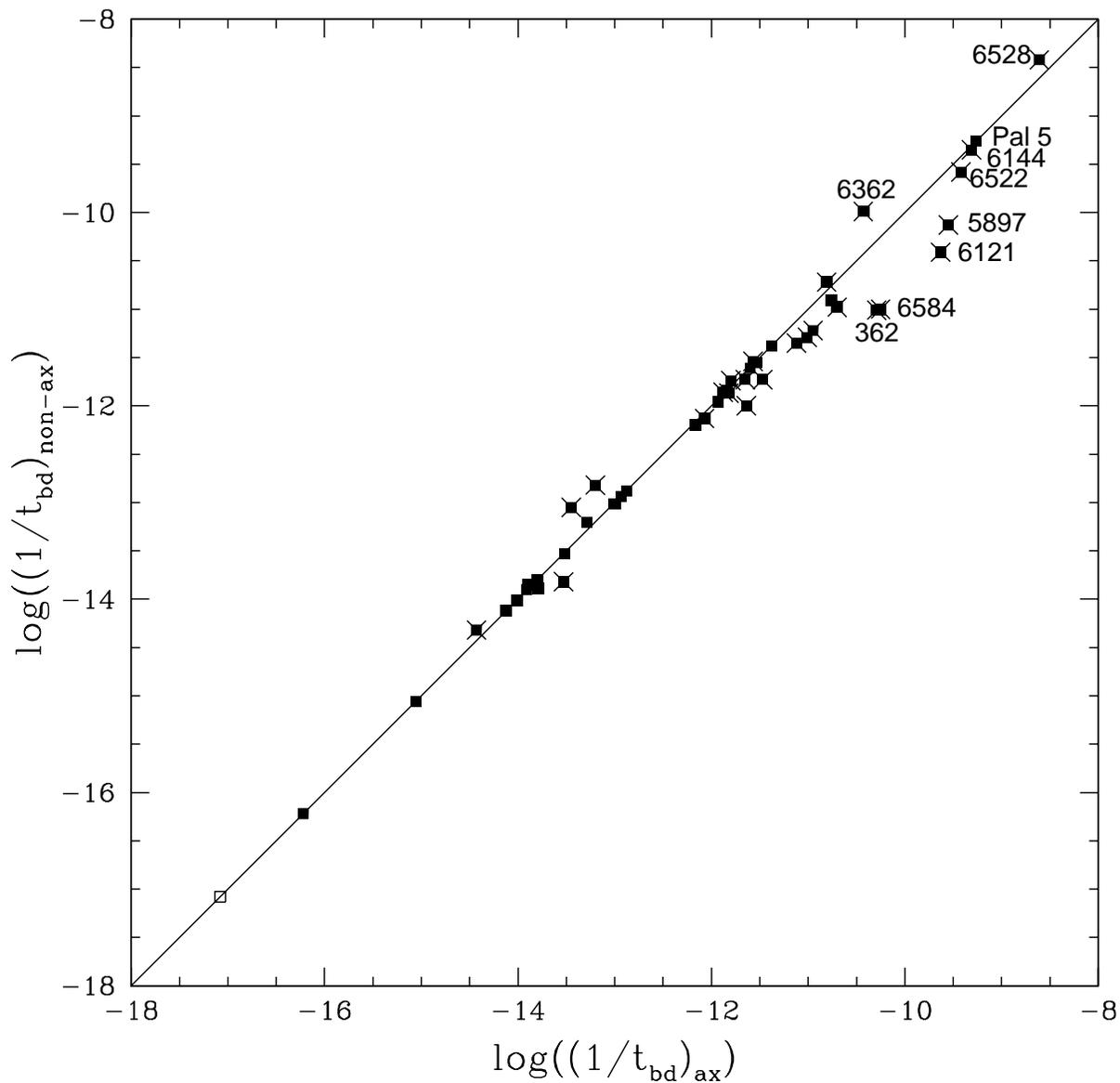}
\caption {Comparison of the total destruction rates due to the bulge and disk,
in the axisymmetric and non-axisymmetric potentials. We plot only the points
corresponding to the 'central' orbits in Figure \ref{fig10}. Some NGC
and Pal numbers are shown. Marked squares correspond to clusters with
$<$$r_{min}$$>$ $<$ 3 kpc.}
\label{fig11}
\end{figure}

In Figures \ref{fig8}, \ref{fig9}, and \ref{fig10} we compare the
destruction rates due to the bulge, the disk, and bulge and disk,
respectively, in the axisymmetric and non-axisymmetric potentials. In
these figures, the empty squares, empty triangles, and filled squares
show the values obtained with the extreme minimum and maximum energy
orbits, and the 'central' orbit, respectively.  In Figure \ref{fig8},
some clusters (NGC 1904, NGC 2298, NGC 4147, NGC 5897, NGC 6093, NGC
6316, NGC 6712, NGC 6934, NGC 7006) have large variations of their
destruction rates, as given by the extreme orbits. For the rest of the
sample the mean deviation from the value given by the 'central' orbit
is around 0.9 dex and 0.78 dex in the axisymmetric and
non-axisymmetric potentials, respectively. The points (shown as filled
squares) with largest deviation from the line of coincidence
correspond to clusters with $<$$r_{min}$$>$ $<$ 3 kpc, $r_{min}$ being
the perigalactic distance; for these clusters the effect of the bar is
important.  In Figure \ref{fig9}, the mean deviation of destruction
rates from that of the 'central' orbit is 0.4 dex in both potentials,
omitting the clusters listed above for which the variation is
large. In this case, there is no important deviation from the line of
coincidence, even for clusters with perigalactic points inside the bar
region. The combined effects of bulge and disk shown in Figure
\ref{fig10}, give a corresponding mean deviation of destruction rates
of 0.6 dex and 0.57 dex in the axisymmetric and non-axisymmetric
potentials, respectively. This figure shows that most clusters have
similar total (bulge+disk) destruction rates in both
potentials. Figure \ref{fig11} shows some detail of Figure
\ref{fig10}. Here we plot only the the points corresponding to the
'central' orbits, adding some NGC and Pal numbers, and showing with
marked squares the clusters with $<$$r_{min}$$>$ $<$ 3 kpc.  In NGC
362, NGC 5897, NGC 6121, and NGC 6584, the total destruction rate is
sensibly smaller for the non-axisymmetric potential; considering only
the effects of tidal shocks with the bulge and disk, in the barred
potential five clusters are expected to be destroyed during the next
Hubble time (taken as $10^{10}$ yr): Pal 5, NGC 6144, NGC 6362, NGC
6522, and NGC 6528.  Considering the interval defined by the extreme
minimum and maximum energy orbits, NGC 5897, NGC 6121, and NGC 6584
could also be destroyed during the next Hubble time. In the
axisymmetric potential the six clusters expected to be destroyed by
the tidal shocks are Pal 5, NGC 5897, NGC 6121, NGC 6144, NGC 6522,
and NGC 6528. Considering also the extreme orbits, NGC 6093, NGC 6316,
and NGC 6712 could be added to this list. A similar result is obtained
by \citet{D99b}, who find that, in their axisymmetric potential, five
clusters are destroyed in a Hubble time under the combined effect of
the tidal shocks with the bulge and disk (see the bottom left panel in
their figure 8). The total destruction rate will increase if we
include the evaporation by two-body relaxation; the destruction rates
due to this effect are given by G\&O97 in their Table 3 (using the
value 3 for the mass-to-light ratio).

\begin{figure}
\plotone{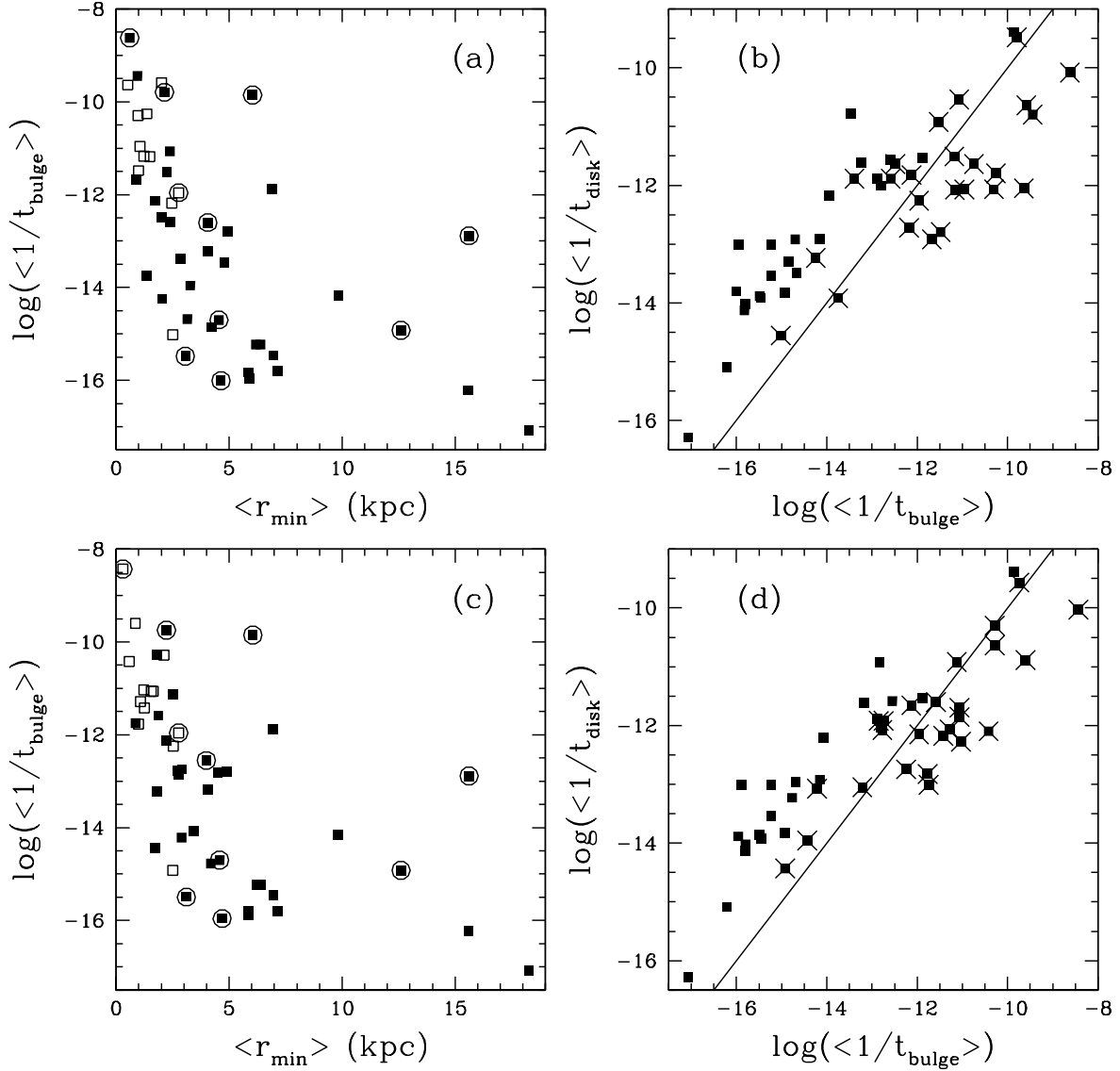}
\caption {Destruction rates with the 'central' orbits. 
We show the values of the total destruction rate due to the bulge as
a function of $<$$r_{min}$$>$, the average minimum distance to the Galactic
center, in (a) the axisymmetric, and (c) the non-axisymmetric potential.
The squares with a circle correspond to clusters with a mass less than
$10^5 M_{\odot}$, and the empty squares correspond to clusters with
$<$$r_{min}$$>$ $<$ 3 kpc and orbital eccentricity $e > 0.6$.
In (b) and (d) we give the comparison of the destruction rates due to the
bulge and disk, in the axisymmetric and non-axisymmetric potentials,
respectively. Marked squares correspond to clusters with
$<$$r_{min}$$>$ $<$ 3 kpc; the line of coincidence is plotted in both frames.}
\label{fig12}
\end{figure}

In Figure \ref{fig12} we plot only the values obtained with the
'central' orbits. We show how the destruction rates due to the bulge
depend on the mean perigalactic distance, in both the axisymmetric
(frame (a)) and non-axisymmetric (frame (c)) potentials. Clusters with
perigalactic points close to the Galactic center and large orbital
eccentricities (empty squares) have, in general, greater bulge
destruction rates. In these frames, points corresponding to clusters
with a mass less than $10^5 M_{\odot}$ are shown as circles. Frames
(b) and (d) of Figure \ref{fig12} give the comparison of the
destruction rates due to the bulge and disk, in the axisymmetric and
non-axisymmetric potentials, respectively. These frames show that
bulge shocking dominates in the bar region, as found by \citet{AHO88}.

Figures \ref{fig8} to \ref{fig12} show that the destruction rates are
very similar in both the axisymmetric and non-axisymmetric Galactic
potentials. A similar result was obtained by \citet{LOA92}.  However,
our destruction rates are sensibly smaller than those obtained by
G\&O97, who find a much larger destruction rate due to the Galactic
bulge and disk. A similar conclusion was reached by \citet{D99b} when
comparing their results with those of G\&O97, employing an
axisymmetric potential. Our approach to compute the Galactic orbits of
globular clusters is the same as that considered by \citet{D99b}, that
is, the use of the available data on absolute proper motions and
radial velocities.  On the other hand, G\&O97 consider a statistical
procedure to compute the orbits. In their study, \citet{D99b} have
computed the destruction rates due to tidal shocks at the half-mass
radius, while we have taken an average over the whole cluster,
including the adiabatic corrections of G\&O99.  However, this
improvement in the computation of destruction rates cannot account for
the difference with G\&O97 results. Thus, it would be of great
interest to use the rigorous Fokker-Planck approach of G\&O97 along
with full space motion data, and galactic orbits, in order to achieve
a better estimate of the destruction rates.

\section{CONCLUSIONS}\label{concl}

We have obtained orbits for 48 globular clusters in both a barred and
an axisymmetric galactic potential. The orbits of outer clusters
(those with pericentric distances greater than about 4 kpc) are
largely unaffected by the bar. The largest changes were found to occur
for 'inner' clusters, i.e., those whose orbits reside mostly within
the bar region. The main changes that the bar causes in the orbits are
larger vertical and radial excursions. In general
the bar causes no net global changes in the energy or the z-component
of the angular momentum, computed in an inertial frame. However, there
are cases where jumps in these quantities do occur, even causing a
temporary reversal of the sense of rotation of the orbit. Tidal radii
have been computed with a new expression, and with a numerical
evaluation of the relevant quantities along the orbit. No noticeable
changes due to the bar were found for most of the clusters. When
changes do occur, they generally make the computed tidal radii
somewhat larger in the presence of a bar. We find that the destruction
rates due to shocks with the galactic bulge and disk are not strongly
affected if we consider a barred galactic potential; in some cases
(NGC 362, NGC 5897, NGC 6121, and NGC 6584) the total destruction rate
due to both effects can even $decrease$ when using the barred
potential.  We thus concur with the result of \citet{LOA92}, that the
similarity of destruction rates in axisymmetric and barred potentials
does not rule out the existence of a galactic bar. Although we have
included adiabatic corrections in the computation of tidal-shock
destruction rates, these rates are smaller than those obtained by
G\&O97. Thus, to clarify this difference, the absolute proper motion
data should be used with the G\&O97 approach.

\appendix
\section{Alternative Formulation for Estimating the Tidal Radius}
%\begin{center}
%\appendix{APPENDIX A}\label{apendA}
%\end{center}

For the motion of a globular cluster of mass $M_c$ in a galactic force
field, on an orbit of eccentricity $e$, an approximation to the tidal
radius of the cluster evaluated at a perigalactic point with
galactocentric distance $r_{min}$, is given by King's (1962) formula
$r_K = [M_c/M_g(3+e)]^{1/3}r_{min}$.  This formula needs an
approximate value of the 'effective' galactic mass $M_g$, which can be
taken as the point mass that produces the actual force felt by the
cluster at the perigalactic point \citep{AM88}. An alternative
approximate formula for the tidal radius of a cluster in a general
galactic field can be obtained, requiring only that at certain
position on the line joining the cluster and the galactic center, a
test particle with velocity {\boldmath $v'$} = 0, measured in a
non-inertial reference frame at the center of the cluster, has an
acceleration {\boldmath $a'$} in this frame with no component along
this line.

Take a galactic inertial frame with its origin at the galactic center;
Cartesian axes in this frame are $x, y, z$, the $z$- axis being
perpendicular to the galactic plane. A cluster moving in the galactic
field has angular spherical coordinates ($\varphi,\theta$) measured in
the inertial frame. The primed non-inertial frame at the cluster has
its Cartesian $x'$- axis pointing always toward the galactic center,
and the $y'$- axis pointing in the opposite direction of the unit
vector in the azimuthal direction, {\boldmath $e$}$_{\varphi}$. The
angular velocity of the primed frame with respect to the inertial
frame is {\boldmath ${\Omega}$} = $\dot{\varphi}${\boldmath $k$} +
$\dot{\theta}${\boldmath $e$}$_{\varphi}$ =
$-\dot{\varphi}\cos{\theta}${\boldmath $i'$}$-\dot{\theta}${\boldmath
$j'$}+ $\dot{\varphi}\sin{\theta}${\boldmath $k'$}; with {\boldmath
$k$}, {\boldmath $i'$}, {\boldmath $j'$}, {\boldmath $k'$} the unit
vectors on the $z, x', y', z'$- axes, respectively. The position of
the cluster in the inertial frame is {\boldmath $r$}$_c$ and any point
in space has a position {\boldmath $r$}, {\boldmath $r'$} in the
inertial and primed frames, respectively. The acceleration {\boldmath
$a'$} of a test particle at {\boldmath $r'$} is given by (see e.g.,
\citet{SY71}) {\boldmath $a'$} = {\boldmath $a$}$-${\boldmath
$a$}$_0$$-$ 2{\boldmath ${\Omega}$}$\times${\boldmath $v'$}$-$
{\boldmath ${\Omega}$}$\times$({\boldmath
${\Omega}$}$\times${\boldmath $r'$})$-$ ($\frac{d}{dt}${\boldmath
${\Omega}$})$\times${\boldmath $r'$}, with {\boldmath $a$} the
acceleration at {\boldmath $r'$} measured in the inertial frame, and
{\boldmath $a$}$_0$ the acceleration of the primed frame with respect
to the inertial frame.  These accelerations have the expressions
{\boldmath $a$} =$-GM_c${\boldmath $r'$}/$|${\boldmath $r'$}$|$$^3$
$-(\nabla{\Phi})_{\bf r}$, {\boldmath $a$}$_0$ = $-(\nabla{\Phi})_{\bf
r_c}$; $\Phi$ is the galactic potential per unit mass and the galactic
acceleration $-\nabla{\Phi}$ appears evaluated at the position
{\boldmath $r$} of the test particle and at the position {\boldmath
$r$}$_c$ of the cluster.  We are assuming that {\boldmath $r'$} lies
outside the spherical mass distribution of the cluster. Now we take
{\boldmath $r'$} on the positive $x'$- axis and find the instantaneous
position $x'$ at which $\ddot{x}'$ = 0 with {\boldmath $v'$} = 0. The
term ($\frac{d}{dt}${\boldmath ${\Omega}$})$\times${\boldmath $r'$}
has no $x'$- component, and with {\boldmath ${\Omega}$} given above
the condition satisfied by $x'$ is

\begin{equation} -\frac{GM_c}{x'^2}-\mbox{\boldmath $i'$}\cdot \left [
(\nabla{\Phi})_{\bf r} -(\nabla{\Phi})_{\bf r_c} \right ] +
x'(\dot{\theta}^2 + \dot{\varphi}^2\sin^2{\theta}) = 0.
\label{galL1} \end{equation}

We call $F_{x'}$ =$-${\boldmath $i'$}$\cdot \nabla{\Phi}$, the $x'$-
component of the galactic acceleration along the $x'$- axis. Then
$-${\boldmath $i'$}$\cdot [(\nabla{\Phi})_{\bf r} -(\nabla{\Phi})_{\bf
r_c}] \simeq x'(\partial F_{x'}/\partial x')_{{\bf r'}= 0}$ and the
required solution of Eq. (\ref{galL1}) is approximately

\begin{equation} r_{\ast} \equiv x' = \left [ \frac{GM_c}{\left (\frac{\partial
F_{x'}}
{\partial x'} \right )_{{\bf r'}= 0}+ \dot{\theta}^2 + \dot{\varphi}^2\sin^2
{\theta}} \right ]^{1/3}. \label{x'gal} \end{equation}

Eq. (\ref{x'gal}), which has the form of equation (7) in King's (1962)
analysis, can also be obtained if we start with Cartesian axes $x',
y', z'$ as follows: take the $x'$- axis as before but now the $z'$-
axis points in the direction {\boldmath $k'$} = $(${\boldmath
$r$}$_c$$\times${\boldmath $v$}$_c$$) /$$|${\boldmath
$r$}$_c$$\times${\boldmath $v$}$_c$$|$, with {\boldmath $r$}$_c$,
{\boldmath $v$}$_c$ the instantaneous position and velocity of the
cluster in the inertial frame. The instantaneous angular velocity of
the new primed frame is {\boldmath ${\Omega}$} = {\boldmath
$k'$}$|${\boldmath $v$}$_{ct}$$|$/$|${\boldmath $r$}$_c$$|$, with
{\boldmath $v$}$_{ct}$ = $|${\boldmath $r$}$_c$$|$$\dot{\varphi}
\sin{\theta}${\boldmath $e$}$_{\varphi}$ + $|${\boldmath
$r$}$_c$$|$$\dot {\theta}${\boldmath $e$}$_{\theta}$ the tangential
velocity of the cluster.  Then, with the steps followed above, we
arrive again at Eq. (\ref{x'gal}).

We can apply Eq. (\ref{x'gal}) at any point of the galactic orbit of
the cluster, and in particular at a perigalactic point the resulting
tidal radius $r_{\ast}$ can be compared with that obtained using
King's (1962) formula, $r_K$.

%\begin{center}
%\appendix{APPENDIX B}\label{apendB}
%\end{center}
\section{Simplification of the ${\chi}(|${\boldmath $r$}$_p|)$ Factor}

The analysis made by \citet{S58} (see also \citet{S87}) for the change
of velocity of a particle due to the tidal force arising in the
gravitational interaction of two concentrated masses (with the
particle bounded to one of these masses), can be easily extended to
obtain the change of velocity of a star in a cluster passing by a
bulge with a spherically symmetric mass density ${\rho}_b(r)$, with
$r$ galactocentric distance. The cluster reaches a pericenter position
{\boldmath $r$}$_p$ with a velocity {\boldmath $v$}$_p$ with respect
to a galactic inertial frame. In the impulse approximation the cluster
is supposed to move on the line passing at {\boldmath $r$}$_p$ with
constant velocity {\boldmath $v$}$_p$. An inertial Cartesian frame is
defined with its center at the cluster; the $x$- axis is parallel to
{\boldmath $r$}$_p$, pointing in the $-${\boldmath $r$}$_p$ direction,
and the $y$- axis points along the line of motion. A star in the
cluster has coordinates $x, y, z$. The total changes of the star's
velocity components in this inertial frame are

\begin{equation} {\Delta}v_x = 2GM_bx \int_{0}^{\infty} \frac{{\mu}_b(\xi(t))}
{{\xi}^3(t)} \left [2- \frac{3|\mbox{\boldmath $v$}_p|^2 t^2}{{\xi}^2(t)}
\right ]dt + 8{\pi}Gx \int_{0}^{\infty} {\rho}_b(\xi(t)) \left [
\frac{|\mbox{\boldmath $v$}_p|^2 t^2}{{\xi}^2(t)} - 1 \right ]dt,
\label{dvx} \end{equation}

\begin{equation} {\Delta}v_y = 2GM_by \int_{0}^{\infty} \frac{{\mu}_b(\xi(t))}
{{\xi}^3(t)} \left [\frac{3|\mbox{\boldmath $v$}_p|^2 t^2}{{\xi}^2(t)} -1
\right ]dt - 8{\pi}Gy \int_{0}^{\infty} {\rho}_b(\xi(t))
\frac{|\mbox{\boldmath $v$}_p|^2 t^2}{{\xi}^2(t)}dt,
\label{dvy} \end{equation}

\begin{equation} {\Delta}v_z = -2GM_bz \int_{0}^{\infty} \frac{{\mu}_b(\xi(t))}
{{\xi}^3(t)}dt, \label{dvz} \end{equation}

\noindent with ${\mu}_b(r) = M_b(r)/M_b$, $M_b(r)$ is the mass of the
bulge within a sphere of galactocentric radius $r$, $M_b$ is the total
mass of the bulge, and $\xi(t) = (|${\boldmath $r$}$_p|^2 +
|${\boldmath $v$}$_p|^2 t^2)^ {1/2}$. In the case considered by
\citet{S58} we have ${\mu}_b(r)$ = 1, ${\rho}_b(r)$ = 0 and
Eqs. (\ref{dvx}), (\ref{dvy}), (\ref{dvz}) give his equation (8) (a
minus sign is missing in his expression of ${\Delta}v_z$; correct in
equation (5.35) of \citet{S87}).

Now we change the integrations to the variable $\xi$, and a second
change to the variable $u \equiv \xi/|${\boldmath $r$}$_p|$. Also, we
use $d{\mu}_b/dr = 4{\pi}r^2{\rho}_b(r)/M_b$, and following
\citet{G99a} define $\acute{{\mu}_b}(r) \equiv r(d{\mu}_b/dr) =
d{\mu}_b/d{\ln}r$. Thus, Eq.  (\ref{dvy}) can be written as

\begin{eqnarray} {\Delta}v_y &  = &  \frac{2GM_by}{|\mbox{\boldmath $v$}_p|
|\mbox{\boldmath $r$}_p|^2}  \left \{ \int_{1}^{\infty}
\frac{{\mu}_b(u|\mbox{\boldmath $r$}_p|)}{u^2} \left (2 - \frac{3}{u^2}
\right ) \frac{du}{(u^2-1)^{1/2}} + \right. \nonumber \\
& &  \left. + \int_{1}^{\infty}
\frac{\acute{\mu}_b(u|\mbox{\boldmath $r$}_p|)}{u^2} \left (\frac{1}{u^2}
-1 \right ) \frac{du}{(u^2-1)^{1/2}} \right \} \label{dvyy}.
\end{eqnarray}

\noindent Again, following \citet{G99a} we define

\begin{equation} I_0(|\mbox{\boldmath $r$}_p|) = \int_{1}^{\infty}
{\mu}_b(u|\mbox{\boldmath $r$}_p|) \frac{du}{u^2(u^2-1)^{1/2}},
\label{i0} \end{equation}

\begin{equation} I_1(|\mbox{\boldmath $r$}_p|) = \int_{1}^{\infty}
\acute{\mu}_b(u|\mbox{\boldmath $r$}_p|) \frac{du}{u^2(u^2-1)^{1/2}},
\label{i1} \end{equation}

\begin{equation} J_0(|\mbox{\boldmath $r$}_p|) = \int_{1}^{\infty}
{\mu}_b(u|\mbox{\boldmath $r$}_p|) \frac{du}{u^4(u^2-1)^{1/2}},
\label{j0} \end{equation}

\begin{equation} J_1(|\mbox{\boldmath $r$}_p|) = \int_{1}^{\infty}
\acute{\mu}_b(u|\mbox{\boldmath $r$}_p|) \frac{du}{u^4(u^2-1)^{1/2}},
\label{j1} \end{equation}

\noindent and Eq. (\ref{dvyy}) is

\begin{equation} {\Delta}v_y = \frac{2GM_by}{|\mbox{\boldmath $v$}_p|
|\mbox{\boldmath $r$}_p|^2} \left [ 2I_0(|\mbox{\boldmath $r$}_p|)
- 3J_0(|\mbox{\boldmath $r$}_p|) - I_1(|\mbox{\boldmath $r$}_p|) +
J_1(|\mbox{\boldmath $r$}_p|)\right ].
\label{dvyyy} \end{equation} 

\noindent In the same way, Eqs. (\ref{dvx}), (\ref{dvz}) can be written as

\begin{equation} {\Delta}v_x = \frac{2GM_bx}{|\mbox{\boldmath $v$}_p|
|\mbox{\boldmath $r$}_p|^2} \left [- I_0(|\mbox{\boldmath $r$}_p|)
+ 3J_0(|\mbox{\boldmath $r$}_p|) - J_1(|\mbox{\boldmath $r$}_p|) \right ],
\label{dvxxx} \end{equation}
 
\begin{equation} {\Delta}v_z = - \frac{2GM_bz}{|\mbox{\boldmath $v$}_p|
|\mbox{\boldmath $r$}_p|^2} I_0(|\mbox{\boldmath $r$}_p|).
\label{dvzzz} \end{equation}

\noindent Eqs. (\ref{dvyyy}), (\ref{dvxxx}), (\ref{dvzzz}) are the changes of
the velocity components obtained by \citet{G99a} (see their equation
(10)).

\noindent Eqs. (\ref{dvyyy}), (\ref{dvxxx}) can be reduced as follows. An 
integration by parts shows that (assuming, as above, that
$|${\boldmath $r$}$_p| \neq 0$)

\begin{equation} \int_{0}^{\infty} \frac{{\mu}_b(\xi(t))}{{\xi}^3(t)}dt =
3|\mbox{\boldmath $v$}_p|^2 \int_{0}^{\infty} \frac{{\mu}_b(\xi(t))t^2}
{{\xi}^5(t)}dt - \frac{4{\pi}|\mbox{\boldmath $v$}_p|^2}{M_b} \int_{0}^{\infty}
\frac{{\rho}_b(\xi(t))t^2}{{\xi}^2(t)}dt. \label{intp} \end{equation}

\noindent Thus, returning to Eq. (\ref{dvy}) we find the same result
as that obtained when the bulge is a point mass, namely
 
\begin{equation} {\Delta}v_y = 0, \label{dvyf} \end{equation}

\noindent whereby  $2I_0(|${\boldmath $r$}$_p|)
- 3J_0(|${\boldmath $r$}$_p|) - I_1(|${\boldmath $r$}$_p|) +
J_1(|${\boldmath $r$}$_p|) = 0$, from Eq. (\ref{dvyyy}). Thus,
$3J_0(|${\boldmath $r$}$_p|) - I_0(|${\boldmath $r$}$_p|)
- J_1(|${\boldmath $r$}$_p|) = I_0(|${\boldmath $r$}$_p|) -
I_1(|${\boldmath $r$}$_p|)$, and Eq. (\ref{dvxxx}) reduces to

\begin{equation} {\Delta}v_x = \frac{2GM_bx}{|\mbox{\boldmath $v$}_p|
|\mbox{\boldmath $r$}_p|^2} \left [ I_0(|\mbox{\boldmath $r$}_p|)
- I_1(|\mbox{\boldmath $r$}_p|) \right ].
\label{dvxf} \end{equation}

\noindent Then only two integrals, $I_0(|${\boldmath $r$}$_p|)$ and
$I_1(|${\boldmath $r$}$_p|)$, are needed to compute the change of the
velocity components.

\noindent There is another way to write Eq. (\ref{dvxf}). Integrating
by parts we find

\begin{equation} I_0(|\mbox{\boldmath $r$}_p|) = 1 - \frac{4{\pi}
|\mbox{\boldmath $r$}_p|^3}{M_b} \int_{1}^{\infty}{\rho}_b
(u|\mbox{\boldmath $r$}_p|)(u^2-1)^{1/2}udu,
\label{i00} \end{equation}

\noindent and $I_1(|${\boldmath $r$}$_p|)$ can be written as

\begin{equation} I_1(|\mbox{\boldmath $r$}_p|) = \frac{4{\pi}
|\mbox{\boldmath $r$}_p|^3}{M_b} \int_{1}^{\infty} \frac{{\rho}_b
(u|\mbox{\boldmath $r$}_p|)udu}{(u^2-1)^{1/2}},
\label{i11} \end{equation}

\noindent thus

\begin{equation} {\Delta}v_x = \frac{2GM_bx}{|\mbox{\boldmath $v$}_p|
|\mbox{\boldmath $r$}_p|^2} \left [1 - \frac{4{\pi}|\mbox{\boldmath $r$}_p|^3}
{M_b}{\Im}_2(|\mbox{\boldmath $r$}_p|) \right ],
\label{dvxff} \end{equation}

\noindent with

\begin{equation} {\Im}_2(|\mbox{\boldmath $r$}_p|) =
\int_{1}^{\infty}\frac{{\rho}_b(u|\mbox{\boldmath $r$}_p|)u^3du}
{(u^2-1)^{1/2}}. \label{integ2} \end{equation}

\noindent In Section \ref{bulbo} we call
${\Im}_1(|${\boldmath $r$}$_p|)$ = $I_0(|${\boldmath $r$}$_p|)$.

%%%%%%%%%%%%%%%%%%%%%%%%%%%%%%%%%%%%%%%%%%%%%%%%%%%%%%%%%%%%%%%%%%%%%%%%%%

%\clearpage

\clearpage
\begin{deluxetable}{ccccccccccccc}
\tabletypesize{\scriptsize}
\rotate
\tablecaption{Orbital properties with the axisymmetric potential \label{tbl-1}}
\tablewidth{0pt}
\tablehead{
\colhead{Cluster} & \colhead{$(r_{min})_{min}$} & \colhead{$<$$r_{min}$$>$} &
\colhead{$(r_{max})_{max}$} & \colhead{$<$$r_{max}$$>$} & \colhead{($|z|_{max})
_{max}$} & \colhead{$<$$|z|_{max}$$>$} & \colhead{$<$$e$$>$} & \colhead{$E$} &
\colhead{$h$} & \colhead{$r_{K}$} & \colhead{$r_{\ast}$} &
\colhead{$r_{L}$} \\ \colhead{} & \colhead{($kpc$)} & \colhead{($kpc$)} &
\colhead{($kpc$)} & \colhead{($kpc$)} & \colhead{($kpc$)} &
\colhead{($kpc$)} & \colhead{} & \colhead{(10$kms^{-1})^2$} &
\colhead{(10$kms^{-1}kpc$)} & \colhead{($pc$)} & \colhead{($pc$)} &
\colhead{($pc$)}
}
\startdata
NGC 104 & 5.81 & 5.94 & 7.90 & 7.82 & 3.51 & 3.19 & 0.137 & $-1267.51$ &
121.39 & 100.1 & 113.6 & 56.1 \\
       & 4.96 & 5.10 & 7.84 & 7.79 & 3.17 & 2.76 & 0.209 & $-1297.54$ &
113.12 & 90.5 & 101.5 & \\
       & 6.79 & 6.91 & 7.98 & 7.90 & 3.83 & 3.65 & 0.067 & $-1234.49$ &
129.82 & 112.0 & 128.1 & \\
NGC 288 & 3.51 & 4.05 & 12.39 & 12.29 & 9.87 & 6.91 & 0.506 & $-1160.77$ &
$-66.87$ & 34.1 & 34.5 & 33.1 \\
         & 1.45 & 2.16 & 12.16 & 11.91 & 9.45 & 5.59 & 0.695 & $-1206.96$ &
$-38.65$ & 22.3 & 21.1 & \\
         & 6.03 & 6.41 & 13.02 & 12.86 & 10.56 & 8.12 & 0.335 & $-1101.99$ &
$-95.17$ & 46.1 & 49.1 & \\
NGC 362 & 0.50 & 0.78 & 12.39 & 11.95 & 7.09 & 1.51 & 0.878 & $-1248.36$ &
$-27.30$ & 18.4 & 17.2 & 39.8 \\
        & 0.54 & 1.11 & 10.86 & 10.10 & 6.08 & 2.27 & 0.803 & $-1309.71$ &
$-28.07$ & 18.7 & 18.1 & \\
        & 0.54 & 1.40 & 14.17 & 13.60 & 11.37 & 5.76 & 0.816 & $-1164.84$ &
$-21.78$ & 21.4 & 19.0 & \\
NGC 1851 & 5.76 & 5.88 & 37.53 & 37.50 & 15.18 & 8.50 & 0.729 & $-750.45$ &
207.16 & 67.7 & 64.2 & 41.2 \\
         & 4.59 & 4.73 & 26.31 & 26.28 & 10.05 & 5.83 & 0.695 & $-888.71$ &
162.17 & 59.5 & 56.9 & \\
         & 5.99 & 6.08 & 58.62 & 58.58 & 25.03 & 12.67 & 0.812 & $-584.39$ &
230.93 & 69.1 & 62.8 & \\
NGC 1904 & 4.06 & 4.29 & 21.58 & 21.54 & 9.95 & 6.02 & 0.668 & $-965.01$ &
133.40 & 48.2 & 46.3 & 31.3 \\
         & 1.57 & 1.88 & 19.49 & 19.40 & 8.21 & 3.91 & 0.823 & $-1028.90$ &
66.85 & 26.0 & 23.0 & \\
         & 9.38 & 9.52 & 24.18 & 24.11 & 14.20 & 9.93 & 0.434 & $-873.68$ &
225.95 & 84.8 & 87.5 & \\
NGC 2298 & 4.14 & 4.63 & 17.87 & 17.77 & 14.14 & 9.19 & 0.587 & $-1024.63$ &
$-85.22$ & 31.7 & 31.3 & 20.2 \\
         & 0.46 & 1.38 & 16.29 & 15.24 & 12.50 & 5.17 & 0.836 & $-1123.79$ &
$-25.23$ & 12.3 & 11.2 & \\
         & 10.15 & 10.36 & 23.52 & 23.42 & 18.51 & 13.45 & 0.386 & $-872.77$ &
$-176.66$ & 55.7 & 58.2 & \\
Pal 3 & \nodata & \nodata & \nodata & \nodata & \nodata & \nodata & \nodata &
\nodata & \nodata & \nodata & \nodata & 129.7 \\
      & 81.92 & 81.92 & 128.27 & 128.26 & 116.90 & 99.24 & 0.220 & $-190.11$ &
742.66 & 197.3 & 218.4 & \\
      & \nodata & \nodata & \nodata & \nodata & \nodata & \nodata & \nodata &
\nodata & \nodata & \nodata & \nodata & \\
NGC 4147 & 1.88 & 2.70 & 30.68 & 30.65 & 26.79 & 15.25 & 0.838 & $-838.05$ &
42.49 & 21.6 & 18.2 & 35.4 \\
         & 2.06 & 2.91 & 23.59 & 23.55 & 23.54 & 14.24 & 0.782 & $-934.95$ &
5.81 & 22.1 & 20.3 & \\
         & 9.50 & 9.73 & 46.96 & 46.89 & 41.71 & 25.29 & 0.656 & $-651.33$ &
154.86 & 51.1 & 49.0 & \\
NGC 4590 & 9.72 & 9.82 & 32.30 & 32.24 & 18.52 & 11.97 & 0.533 & $-778.60$ &
257.23 & 73.2 & 73.4 & 90.0 \\
         & 9.06 & 9.19 & 20.61 & 20.54 & 11.09 &  8.05 & 0.382 & $-926.50$ &
218.60 & 70.7 & 74.1 & \\
         & 10.47 & 10.59 & 52.67 & 52.63 & 32.55 & 19.45 & 0.665 & $-608.35$ &
298.93 & 76.0 & 73.1 & \\
NGC 5024 & 15.41 & 15.57 & 37.80 & 37.73 & 35.92 & 25.09 & 0.416 & $-692.02$ &
136.30 & 157.8 & 162.5 & 112.6 \\
         & 10.60 & 10.87 & 19.64 & 19.49 & 18.31 & 14.28 & 0.284 & $-918.32$ &
98.39 & 125.0 & 133.8 & \\
         & 17.64 & 17.73 & 108.76 & 108.72 & 105.52 & 60.17 & 0.720 &
$-336.37$ & 161.00 & 168.3 & 156.0 & \\
NGC 5139 & 1.41 & 1.61 & 6.79 & 6.43 & 3.06 & 1.47 & 0.600 & $-1488.49$ &
$-48.70$ & 56.8 & 54.3 & 87.9 \\
         & 0.91 & 1.34 & 6.94 & 6.27 & 3.40 & 1.65 & 0.648 & $-1502.10$ &
$-37.34$ & 52.6 & 51.5 & \\
         & 1.95 & 1.99 & 6.78 & 6.65 & 1.57 & 1.34 & 0.539 & $-1468.85$ &
$-60.25$ & 70.7 & 69.6 & \\
NGC 5272 & 4.32 & 4.87 & 14.94 & 14.81 & 13.28 & 9.21 & 0.506 & $-1081.68$ &
61.43 & 76.1 & 77.3 & 115.5 \\
         & 3.28 & 3.92 & 13.26 & 13.14 & 12.21 & 8.44 & 0.542 & $-1136.15$ &
39.79 & 64.8 & 64.9 & \\
         & 5.65 & 6.10 & 17.46 & 17.32 & 15.23 & 10.58 & 0.480 & $-1013.35$ &
83.80 & 89.0 & 90.9 & \\
NGC 5466 & 6.56 & 6.93 & 71.46 & 71.40 & 70.59 & 40.67 & 0.823 & $-507.66$ &
$-43.16$ & 50.1 & 44.6 & 158.4 \\
         & 3.92 & 4.51 & 37.41 & 37.30 & 35.13 & 20.64 & 0.784 & $-755.72$ &
$-49.72$ & 37.5 & 33.9 & \\
         & 9.54 & 9.72 & 221.50 & 221.47 & 211.42 & 128.57 & 0.916 & $-172.69$ &
$-44.86$ & 31.6 & 53.4 & \\
Pal 5 & 5.63 & 6.09 & 19.12 & 18.99 & 17.38 & 11.78 & 0.514 & $-982.76$ &
73.29 & 27.2 & 27.6 & 109.9 \\
      & 2.52 & 3.29 & 16.39 & 16.32 & 15.24 &  9.69 & 0.666 & $-1068.39$ &
34.34 & 17.9 & 17.1 & \\
      & 9.81 & 10.08 & 22.99 & 22.87 & 20.49 & 14.77 & 0.388 & $-881.76$ &
124.74 & 39.3 & 41.0 & \\
NGC 5897 & 1.12 & 1.97 & 9.41 & 9.33 & 8.05 & 5.11 & 0.656 & $-1298.83$ &
25.39 & 23.8 & 22.7 & 43.5 \\ 
         & 0.04 & 0.57 & 8.56 & 7.51 & 6.62 & 2.25 & 0.863 & $-1430.58$ & 2.46 &
9.4 & 8.0 & \\
         & 3.47 & 4.05 & 14.35 & 14.25 & 12.06 & 8.11 & 0.558 & $-1108.43$ &
61.80 & 39.1 & 38.3 & \\
NGC 5904 & 1.64 & 2.51 & 49.62 & 49.60 & 45.38 & 24.88 & 0.905 & $-654.46$ &
33.07 & 42.8 & 35.8 & 62.0 \\
         & 0.93 & 1.90 & 30.07 & 30.03 & 26.72 & 14.41 & 0.883 & $-849.11$ &
23.52 & 35.5 & 30.7 & \\
         & 2.45 & 3.23 & 89.15 & 89.12 & 81.60 & 44.97 & 0.930 & $-429.25$ &
40.05 & 62.7 & 51.7 & \\
NGC 6093 & 2.57 & 2.84 & 3.84 & 3.69 & 3.78 & 3.61 & 0.130 & $-1548.60$ &
10.34 & 43.8 & 46.7 & 38.6 \\
         & 0.02 & 0.54 & 5.04 & 3.97 & 3.61 & 1.69 & 0.766 & $-1679.98$ &
$-0.74$ & 15.8 & 15.4 & \\
         & 3.44 & 3.93 & 6.25 & 6.03 & 5.83 & 5.09 & 0.213 & $-1377.54$ &
29.90 & 52.6 & 57.3 & \\
NGC 6121 & 0.33 & 0.66 & 6.81 & 6.14 & 4.05 & 1.09 & 0.806 & $-1538.14$ &
$-18.19$ & 8.6 & 7.9 & 20.8 \\
         & 0.04 & 0.57 & 6.60 & 5.56 & 4.76 & 1.87 & 0.824 & $-1556.13$ &
$-2.55$ & 10.0 & 9.2 & \\
         & 0.78 & 0.92 & 6.96 & 6.66 & 3.36 & 0.72 & 0.757 & $-1513.82$ &
$-34.84$ & 11.7 & 10.9 & \\
NGC 6144 & 2.10 & 2.13 & 2.64 & 2.62 & 2.33 & 2.32 & 0.102 & $-1687.21$ &
$-20.82$ & 23.7 & 24.3 & 82.2 \\
         & 1.07 & 1.41 & 2.94 & 2.90 & 2.53 & 2.08 & 0.351 & $-1717.06$ &
$-16.62$ & 16.3 & 16.7 & \\
         & 1.46 & 1.88 & 3.83 & 3.79 & 3.42 & 2.81 & 0.343 & $-1604.59$ &
$-19.43$ & 20.1 & 20.5 & \\
NGC 6171 & 1.99 & 2.27 & 3.66 & 3.62 & 2.61 & 2.31 & 0.230 & $-1612.79$ &
35.84 & 26.3 & 27.9 & 32.5 \\
         & 0.96 & 1.35 & 3.63 & 3.55 & 2.78 & 2.11 & 0.454 & $-1667.74$ &
21.62 & 17.4 & 17.6 & \\
         & 3.32 & 3.47 & 4.00 & 3.89 & 2.54 & 2.49 & 0.058 & $-1533.24$ &
53.28 & 35.9 & 40.3 & \\
NGC 6205 & 5.72 & 6.20 & 30.33 & 30.22 & 28.86 & 17.57 & 0.660 & $-822.78$ &
$-60.67$ & 80.6 & 79.0 & 56.4 \\
         & 4.21 & 4.82 & 19.00 & 18.87 & 18.23 & 11.85 & 0.594 & $-998.58$ &
$-39.17$ & 69.7 & 69.9 & \\
         & 6.89 & 7.27 & 52.78 & 52.71 & 49.91 & 28.24 & 0.758 & $-618.62$ &
$-87.81$ & 89.1 & 83.3 & \\
NGC 6218 & 2.02 & 2.39 & 5.66 & 5.65 & 3.45 & 2.68 & 0.407 & $-1480.98$ &
48.45 & 28.1 & 28.9 & 25.1 \\
         & 1.33 & 1.74 & 5.35 & 5.24 & 3.32 & 2.38 & 0.505 & $-1531.98$ &
36.68 & 21.7 & 21.6 & \\
         & 2.72 & 3.05 & 6.37 & 6.32 & 3.74 & 3.02 & 0.350 & $-1420.25$ &
61.48 & 33.7 & 35.5 & \\
NGC 6254 & 3.03 & 3.28 & 5.25 & 5.18 & 2.96 & 2.61 & 0.226 & $-1473.24$ &
62.12 & 37.9 & 41.6 & 27.5 \\
         & 2.07 & 2.38 & 4.77 & 4.76 & 2.81 & 2.31 & 0.335 & $-1539.05$ &
47.30 & 30.4 & 32.0 & \\
         & 4.04 & 4.24 & 6.03 & 5.93 & 3.35 & 3.03 & 0.166 & $-1395.28$ &
78.61 & 44.9 & 50.4 & \\
NGC 6266 & 1.19 & 1.36 & 2.61 & 2.36 & 1.05 & 0.85 & 0.269 & $-1839.13$ &
30.47 & 33.5 & 35.7 & 18.0 \\
         & 0.73 & 0.87 & 1.77 & 1.71 & 1.06 & 0.86 & 0.330 & $-1958.02$ &
19.50 & 23.6 & 24.6 & \\
         & 2.10 & 2.17 & 3.08 & 2.89 & 1.15 & 1.09 & 0.144 & $-1719.60$ &
43.43 & 48.5 & 53.1 & \\
NGC 6304 & 1.67 & 1.70 & 3.29 & 3.25 & 0.62 & 0.57 & 0.315 & $-1745.93$ &
43.89 & 23.0 & 24.1 & 23.1 \\
         & 1.25 & 1.26 & 2.48 & 2.41 & 0.64 & 0.58 & 0.313 & $-1854.56$ &
32.74 & 18.2 & 18.9 & \\
         & 2.12 & 2.15 & 4.18 & 4.16 & 0.62 & 0.57 & 0.319 & $-1649.12$ &
56.35 & 27.3 & 28.9 & \\
NGC 6316 & 0.88 & 0.91 & 3.02 & 2.98 & 1.31 & 0.94 & 0.532 & $-1805.20$ & 
$-22.35$ & 18.3 & 18.8 & 19.0 \\
         & 0.17 & 0.31 & 2.15 & 1.84 & 1.33 & 0.62 & 0.711 & $-2027.97$ &
$-8.62$ & 9.0 & 8.1 & \\
         & 1.78 & 1.89 & 4.61 & 4.47 & 1.87 & 1.42 & 0.405 & $-1614.19$ &
$-43.55$ & 32.8 & 34.6 & \\
NGC 6341 & 0.27 & 1.02 & 11.51 & 10.74 & 6.92 & 2.25 & 0.828 & $-1287.00$ &
16.41 & 21.6 & 20.2 & 36.2 \\
         & 1.20 & 1.32 &  9.86 &  9.68 & 4.30 & 2.49 & 0.760 & $-1327.62$ &
23.71 & 22.7 & 21.6 & \\
         & 0.09 & 0.77 & 12.94 & 12.07 & 10.31 & 2.92 & 0.883 & $-1232.39$ &
5.06 & 19.5 & 17.9 & \\
NGC 6362 & 2.15 & 2.36 & 6.08 & 5.74 & 2.45 & 1.71 & 0.417 & $-1502.69$ &
58.29 & 25.0 & 26.2 & 36.9 \\
         & 1.89 & 2.01 & 5.52 & 5.36 & 2.21 & 1.61 & 0.454 & $-1546.21$ &
48.08 & 21.9 & 22.7 & \\
         & 2.67 & 2.93 & 6.16 & 6.15 & 2.89 & 2.32 & 0.355 & $-1443.96$ &
67.17 & 29.3 & 30.9 & \\
NGC 6397 & 2.75 & 2.94 & 6.98 & 6.64 & 2.64 & 1.90 & 0.387 & $-1428.45$ &
74.27 & 27.4 & 29.2 & 10.6 \\
         & 2.24 & 2.26 & 6.59 & 6.39 & 1.71 & 1.51 & 0.479 & $-1469.13$ &
64.04 & 22.3 & 22.2 & \\
         & 3.35 & 3.55 & 6.99 & 6.97 & 2.85 & 2.31 & 0.326 & $-1384.20$ &
84.89 & 30.4 & 32.7 & \\
NGC 6522 & 0.67 & 0.96 & 2.93 & 2.78 & 1.98 & 1.43 & 0.492 & $-1785.03$ &
19.42 & 15.6 & 15.7 & 37.3 \\
         & 0.04 & 0.24 & 1.87 & 1.54 & 1.41 & 0.78 & 0.731 & $-2092.40$ &
$-1.92$ & 7.7 & 6.9 & \\
         & 1.45 & 1.90 & 6.90 & 6.82 & 4.18 & 2.85 & 0.566 & $-1432.96$ &
42.17 & 26.2 & 25.5 & \\
NGC 6528 & 0.45 & 0.60 & 2.51 & 2.23 & 1.23 & 0.63 & 0.573 & $-1931.17$ &
17.82 & 7.9 & 8.1 & 38.1 \\
         & 0.00 & 0.33 & 1.85 & 1.44 & 1.48 & 0.85 & 0.638 & $-2095.91$ &
0.03 & 6.8 & 7.1 & \\
         & 0.98 & 1.19 & 5.19 & 4.79 & 2.31 & 0.97 & 0.602 & $-1625.27$ &
37.10 & 12.3 & 11.9 & \\
NGC 6553 & 2.50 & 2.50 & 13.72 & 13.72 & 0.78 & 0.54 & 0.691 & $-1177.35$ &
98.58 & 33.8 & 31.0 & 14.2 \\
         & 1.89 & 1.90 &  9.73 &  9.73 & 0.77 & 0.54 & 0.673 & $-1334.85$ &
72.08 & 27.7 & 25.8 & \\
         & 3.10 & 3.10 & 19.12 & 19.12 & 0.95 & 0.57 & 0.721 & $-1032.85$ &
127.12 & 38.5 & 35.4 & \\
NGC 6584 & 0.52 & 0.85 & 15.06 & 14.57 & 9.00 & 1.46 & 0.891 & $-1161.57$ &
29.50 & 19.4 & 17.9 & 36.5 \\
         & 0.77 & 0.97 &  8.90 &  8.54 & 5.50 & 2.34 & 0.796 & $-1381.18$ &
9.76 & 15.0 & 14.2 & \\
         & 1.99 & 2.24 & 26.56 & 26.50 & 17.50 & 8.86 & 0.844 & $-898.59$ &
59.69 & 29.2 & 25.7 & \\
NGC 6626 & 2.00 & 2.02 & 3.14 & 3.11 & 0.65 & 0.62 & 0.213 & $-1733.94$ &
47.45 & 34.1 & 36.8 & 18.4 \\
         & 1.25 & 1.26 & 2.65 & 2.58 & 0.67 & 0.61 & 0.343 & $-1834.98$ &
33.46 & 23.4 & 24.2 & \\
         & 2.99 & 3.01 & 3.66 & 3.64 & 0.74 & 0.72 & 0.094 & $-1630.17$ &
63.95 & 45.0 & 51.9 & \\
NGC 6656 & 3.05 & 3.17 & 10.23 & 10.06 & 3.54 & 2.33 & 0.521 & $-1274.96$ &
91.23 & 48.7 & 49.4 & 27.0 \\
         & 2.29 & 2.35 &  8.54 &  8.47 & 1.76 & 1.40 & 0.566 & $-1367.13$ &
75.02 & 40.1 & 39.4 & \\
         & 3.32 & 3.48 & 12.83 & 12.80 & 4.45 & 2.81 & 0.573 & $-1173.35$ &
108.71 & 51.3 & 51.2 & \\
NGC 6712 & 0.91 & 0.99 & 6.87 & 6.61 & 3.72 & 1.94 & 0.739 & $-1490.43$ &
11.45 & 14.7 & 14.2 & 14.9 \\
         & 0.04 & 0.56 & 6.72 & 5.66 & 4.85 & 1.90 & 0.828 & $-1547.13$ &
$-2.59$ & 10.0 & 9.2 & \\
         & 1.36 & 1.38 & 7.87 & 7.85 & 3.16 & 2.13 & 0.702 & $-1415.39$ &
27.29 & 19.1 & 18.5 & \\
NGC 6723 & 1.96 & 2.01 & 2.60 & 2.57 & 2.59 & 2.56 & 0.122 & $-1690.11$ & 2.61 &
31.4 & 31.8 & 26.6 \\
         & 1.15 & 1.47 & 2.61 & 2.58 & 2.42 & 2.08 & 0.279 & $-1741.99$ &
11.96 & 23.1 & 23.6 & \\
         & 1.85 & 2.39 & 3.63 & 3.42 & 3.55 & 3.13 & 0.183 & $-1598.10$ &
$-10.40$ & 34.8 & 37.7 & \\
NGC 6752 & 4.71 & 4.77 & 5.83 & 5.77 & 1.79 & 1.70 & 0.095 & $-1402.51$ &
100.32 & 51.9 & 60.5 & 64.4 \\
         & 4.19 & 4.20 & 5.73 & 5.55 & 1.94 & 1.83 & 0.138 & $-1431.89$ &
89.33 & 47.7 & 54.7 & \\
         & 5.37 & 5.43 & 5.99 & 5.95 & 1.66 & 1.61 & 0.046 & $-1370.16$ &
111.79 & 56.4 & 66.3 & \\
NGC 6779 & 1.01 & 1.06 & 13.69 & 13.59 & 1.51 & 0.88 & 0.856 & $-1197.37$ &
$-46.73$ & 15.0 & 13.4 & 25.2 \\
         & 0.41 & 0.76 & 12.03 & 11.41 & 6.97 & 1.60 & 0.876 & $-1266.17$ &
$-23.36$ & 11.3 & 10.1 & \\
         & 1.76 & 1.83 & 16.19 & 16.14 & 2.53 & 1.64 & 0.797 & $-1111.58$ &
$-74.69$ & 23.0 & 20.5 & \\
NGC 6809 & 0.94 & 1.69 & 6.50 & 6.44 & 6.08 & 4.15 & 0.593 & $-1436.11$ & 12.01 &
22.9 & 22.5 & 25.1 \\
         & 1.13 & 1.80 & 5.68 & 5.63 & 5.25 & 3.78 & 0.523 & $-1479.90$ & 14.89 &
23.8 & 23.6 & \\
         & 0.71 & 1.54 & 7.60 & 7.52 & 7.38 & 4.83 & 0.669 & $-1380.69$ &
$-1.74$ & 20.1 & 19.2 & \\
NGC 6838 & 4.64 & 4.64 & 7.15 & 7.15 & 0.32 & 0.29 & 0.213 & $-1369.37$ &
121.59 & 26.3 & 30.0 & 10.4 \\
         & 3.94 & 3.94 & 7.14 & 7.14 & 0.35 & 0.31 & 0.289 & $-1396.93$ &
109.31 & 23.9 & 26.3 & \\
         & 5.28 & 5.28 & 7.36 & 7.36 & 0.37 & 0.34 & 0.164 & $-1335.35$ &
133.28 & 28.8 & 33.0 & \\
NGC 6934 & 6.91 & 7.20 & 56.01 & 55.93 & 54.62 & 31.37 & 0.772 & $-597.11$ &
$-59.29$ & 58.7 & 53.9 & 38.2 \\
         & 2.33 & 3.14 & 26.49 & 26.43 & 26.23 & 15.72 & 0.788 & $-890.30$ &
$-12.96$ & 33.8 & 30.1 & \\
         & 9.95 & 10.08 & 211.01 & 210.97 & 200.10 & 114.78 & 0.909 & $-181.02$ &
$-110.29$ & 75.3 & 64.2 & \\
NGC 7006 & 18.25 & 18.28 & 111.35 & 111.34 & 56.13 & 31.57 & 0.718 & $-328.03$ &
600.50 & 122.4 & 114.2 & 76.5 \\
         &  6.84 &  6.93 &  56.53 &  56.51 & 21.67 & 11.62 & 0.782 & $-595.58$ &
262.88 & 61.5 & 57.0 & \\
         & \nodata & \nodata & \nodata & \nodata & \nodata & \nodata & \nodata &
\nodata & \nodata & \nodata & \nodata & \\
NGC 7078 & 6.15 & 6.41 & 11.96 & 11.84 & 7.68 & 6.10 & 0.298 & $-1131.02$ &
124.56 & 97.6 & 105.6 & 64.4 \\
         & 4.00 & 4.29 & 10.51 & 10.44 & 5.77 & 4.31 & 0.418 & $-1220.05$ &
98.49 & 74.4 & 77.8 & \\
         & 8.76 & 8.98 & 14.86 & 14.73 & 10.42 & 8.49 & 0.243 & $-1022.28$ &
156.89 & 124.1 & 134.8 & \\
NGC 7089 & 6.59 & 6.99 & 43.52 & 43.44 & 41.47 & 23.79 & 0.723 & $-689.58$ &
$-75.45$ & 96.4 & 89.8 & 71.8 \\
         & 5.59 & 6.10 & 23.64 & 23.51 & 23.12 & 14.84 & 0.588 & $-910.25$ &
$-38.70$ & 88.2 & 86.9 & \\
         & 8.48 & 8.72 & 90.69 & 90.64 & 84.72 & 47.39 & 0.824 & $-415.82$ &
$-127.73$ & 112.1 & 98.1 & \\
NGC 7099 & 3.62 & 4.08 & 7.66 & 7.51 & 6.15 & 5.03 & 0.297 & $-1317.26$ &
$-56.76$ & 42.5 & 45.7 & 42.7 \\
         & 1.96 & 2.48 & 7.10 & 7.06 & 5.36 & 3.94 & 0.483 & $-1390.93$ &
$-40.88$ & 29.9 & 30.1 & \\
         & 4.99 & 5.41 & 9.40 & 9.21 & 7.81 & 6.43 & 0.260 & $-1219.80$ &
$-70.12$ & 52.1 & 56.9 & \\
Pal 12 & 15.45 & 15.60 & 26.24 & 26.14 & 23.45 & 18.71 & 0.253 & $-794.28$ &
172.52 & 44.0 & 47.4 & 96.8 \\
       & 13.60 & 13.78 & 15.56 & 15.41 & 14.05 & 13.35 & 0.056 & $-939.10$ &
122.35 & 40.9 & 45.7 & \\
       & 17.19 & 17.30 & 46.10 & 46.05 & 40.87 & 27.77 & 0.454 & $-623.07$ &
235.93 & 45.5 & 46.9 & \\
Pal 13 & 12.50 & 12.61 & 95.58 & 95.55 & 69.88 & 40.50 & 0.767 & $-390.18$ &
$-339.94$ & 28.1 & 25.9 & 23.3 \\
       &  9.73 &  9.90 & 57.78 & 57.73 & 43.74 & 25.04 & 0.707 & $-577.84$ &
$-238.62$ & 24.4 & 22.9 & \\
       & 15.17 & 15.22 & 211.62 & 211.60 & 149.51 & 86.54 & 0.866 & $-178.16$ &
$-457.11$ & 32.0 & 28.3 & \\
\enddata
\end{deluxetable}

\clearpage
\begin{deluxetable}{cccccccccccc}
\tabletypesize{\scriptsize}
\rotate
\tablecaption{Orbital properties with the non-axisymmetric potential \label{tbl-2}}
\tablewidth{0pt}
\tablehead{
\colhead{Cluster} & \colhead{$(r_{min})_{min}$} & \colhead{$<$$r_{min}$$>$} &
\colhead{$(r_{max})_{max}$} & \colhead{$<$$r_{max}$$>$} & \colhead{$(|z|_{max})
_{max}$} & \colhead{$<$$|z|_{max}$$>$} & \colhead{$<$$e$$>$} & \colhead{$h_{min}$}
 & \colhead{$h_{max}$} & \colhead{$r_{K}$} & \colhead{$r_{\ast}$} \\
\colhead{} & \colhead{($kpc$)} & \colhead{($kpc$)} &
\colhead{($kpc$)} & \colhead{($kpc$)} & \colhead{($kpc$)} &
\colhead{($kpc$)} & \colhead{} & \colhead{(10$kms^{-1}kpc$)} &
\colhead{(10$kms^{-1}kpc$)} & \colhead{($pc$)} & \colhead{($pc$)}
}
\startdata
NGC 104 & 5.74 & 5.88 & 8.11 & 7.86 & 3.56 & 3.16 & 0.145 & 119.27 & 123.16 &
99.5 & 112.2 \\
        & 4.90 & 5.07 & 8.29 & 7.93 & 3.31 & 2.78 & 0.220 & 111.03 & 116.46 &
89.8 & 100.0 \\
        & 6.72 & 6.87 & 8.11 & 7.99 & 3.89 & 3.68 & 0.076 & 129.25 & 131.10 &
111.6 & 127.1 \\
NGC 288 & 3.50 & 4.08 & 13.63 & 12.86 & 11.48 & 7.33 & 0.519 & $-67.37$ & 
$-60.76$ & 33.1 & 33.5 \\
        & 1.05 & 2.27 & 16.52 & 13.34 & 14.61 & 7.16 & 0.710 & $-40.15$ &
$-17.16$ & 22.6 & 21.3 \\
        & 6.05 & 6.42 & 13.23 & 12.94 & 10.81 & 8.20 & 0.337 & $-95.27$ &
$-94.30$ & 46.1 & 49.3 \\
NGC 362 & 0.52 & 1.50 & 13.14 & 11.99 & 10.74 & 5.31 & 0.781 & $-33.82$ & 
$-15.53$ & 25.3 & 22.2 \\
        & 0.82 & 1.56 & 10.06 &  8.49 &  6.05 & 2.82 & 0.690 & $-44.22$ &
$-27.95$ & 25.0 & 23.0 \\
        & 0.61 & 1.32 & 14.13 & 13.37 &  9.65 & 4.81 & 0.822 & $-28.31$ &
$-20.00$ & 25.6 & 21.6 \\
NGC 1851 & 5.76 & 5.89 & 38.46 & 37.68 & 15.23 & 8.48 & 0.730 & 206.29 &
208.77 & 67.5 & 64.0 \\
         & 4.60 & 4.75 & 27.57 & 26.82 & 10.49 & 5.92 & 0.699 & 161.67 &
165.26 & 59.0 & 56.3 \\
         & 5.97 & 6.07 & 60.31 & 59.58 & 25.57 & 12.85 & 0.815 & 230.82 &
232.75 & 68.6 & 62.4 \\
NGC 1904 & 4.02 & 4.27 & 22.05 & 20.29 & 9.72 & 5.88 & 0.652 & 124.73 & 134.84 &
47.8 & 46.1 \\
         & 1.30 & 1.94 & 27.58 & 20.11 & 9.39 & 3.85 & 0.823 & 51.65 & 90.42 &
27.6 & 24.5 \\
         & 9.37 & 9.52 & 24.27 & 24.14 & 14.24 & 9.94 & 0.434 & 225.77 &
226.22 & 84.5 & 87.2 \\
NGC 2298 & 4.15 & 4.64 & 17.90 & 17.66 & 14.20 & 9.16 & 0.585 & $-86.75$ & 
$-84.74$ & 31.7 & 31.4 \\
         & 0.53 & 1.30 & 17.90 & 15.03 & 12.42 & 4.75 & 0.842 & $-33.48$ &
$-15.54$ & 14.3 & 12.5 \\
         & 10.15 & 10.36 & 23.52 & 23.42 & 18.51 & 13.44 & 0.387 & $-176.75$ &
$-176.59$ & 55.7 & 58.1 \\
Pal 3  & \nodata & \nodata & \nodata & \nodata & \nodata & \nodata & \nodata &
\nodata & \nodata & \nodata & \nodata \\
      & 81.92 & 81.92 & 128.27 & 128.26 & 116.90 & 99.24 & 0.220 & 742.66 &
742.66 & 197.3 & 218.4 \\
      & \nodata & \nodata & \nodata & \nodata & \nodata & \nodata & \nodata &
\nodata & \nodata & \nodata & \nodata \\
NGC 4147 & 2.03 & 2.81 & 38.80 & 31.55 & 31.42 & 16.01 & 0.834 & 32.13 & 57.50 &
21.8 & 18.8 \\
         & 2.13 & 3.01 & 26.53 & 24.17 & 25.77 & 14.57 & 0.780 & 0.66 & 13.28 &
22.9 & 20.9 \\
         & 9.50 & 9.73 & 46.92 & 46.79 & 41.59 & 25.25 & 0.655 & 154.58 &
154.90 & 51.1 & 48.5 \\
NGC 4590 & 9.70 & 9.82 & 32.49 & 32.31 & 18.62 & 11.98 & 0.534 & 257.06 &
257.49 & 73.0 & 73.3 \\
         & 8.98 & 9.14 & 21.32 & 20.86 & 11.29 &  8.13 & 0.390 & 218.43 &
220.35 & 70.5 & 73.7 \\
         & 10.47 & 10.58 & 52.92 & 52.75 & 32.68 & 19.48 & 0.666 & 298.83 &
299.16 & 75.9 & 73.0 \\
NGC 5024 & 15.42 & 15.57 & 37.81 & 37.73 & 35.92 & 25.09 & 0.416 & 136.27 & 
136.33 & 157.8 & 162.5 \\
         & 10.61 & 10.88 & 19.64 & 19.48 & 18.31 & 14.28 & 0.283 &  98.21 &
98.52 & 125.0 & 133.9 \\
         & 17.64 & 17.73 & 108.80 & 108.75 & 105.54 & 57.04 & 0.720 & 160.99 &
161.02 & 168.3 & 156.0 \\ 
NGC 5139 & 1.14 & 1.70 & 8.39 & 6.66 & 3.63 & 1.79 & 0.591 & $-55.83$ &
$-34.28$ & 68.7 & 69.7 \\
         & 0.46 & 1.34 & 8.83 & 6.77 & 5.51 & 2.20 & 0.667 & $-45.37$ &
$-15.24$ & 59.2 & 57.9 \\ 
         & 1.82 & 2.28 & 7.53 & 6.32 & 3.19 & 1.71 & 0.469 & $-71.63$ &
$-53.09$ & 75.3 & 77.9 \\
NGC 5272 & 4.27 & 4.85 & 16.48 & 15.35 & 14.35 & 9.32 & 0.520 & 59.24 & 67.26 &
75.6 & 76.4 \\
         & 3.12 & 3.91 & 15.23 & 13.23 & 13.49 & 8.41 & 0.543 & 26.56 & 48.44 &
65.2 & 65.9 \\
         & 5.64 & 6.09 & 17.47 & 17.14 & 15.17 & 10.45 & 0.476 & 82.35 & 84.10 &
88.7 & 90.7 \\
NGC 5466 & 6.57 & 6.94 & 71.51 & 71.00 & 70.31 & 40.45 & 0.822 & $-43.90$ &
$-43.01$ & 50.2 & 44.7 \\
         & 3.89 & 4.51 & 37.42 & 36.74 & 34.94 & 20.29 & 0.782 & $-51.79$ &
$-49.61$ & 37.5 & 34.1 \\
         & 9.54 & 9.72 & 221.26 & 221.03 & 210.93 & 128.25 & 0.915 &
$-44.96$ & $-44.82$ & 31.7 & 53.5 \\
Pal 5 & 5.65 & 6.12 & 19.17 & 18.70 & 17.42 & 11.64 & 0.507 & 71.33 & 73.71 &
27.3 & 27.7 \\
      & 2.55 & 3.33 & 16.83 & 16.24 & 15.39 &  9.84 & 0.661 & 31.58 & 35.99 &
18.2 & 17.4 \\
      & 9.82 & 10.08 & 22.99 & 22.84 & 20.45 & 14.76 & 0.387 & 124.53 & 124.82 &
39.3 & 41.0 \\
NGC 5897 & 1.33 & 2.26 & 12.37 & 10.48 & 8.92 & 5.63 & 0.647 & 24.46 & 44.22 &
25.5 & 24.2 \\
         & 0.05 & 1.00 &  8.29 &  6.90 & 7.21 & 3.09 & 0.759 & $-15.47$ &
17.60 & 14.0 & 12.1 \\
         & 3.50 & 4.14 & 18.74 & 15.72 & 14.25 & 8.56 & 0.582 & 60.98 & 77.85 &
39.1 & 38.4 \\
NGC 5904 & 1.71 & 2.51 & 55.28 & 51.05 & 47.74 & 25.95 & 0.906 & 28.18 & 39.95 &
44.5 & 37.0 \\
         & 0.54 & 1.66 & 31.80 & 26.46 & 26.74 & 13.91 & 0.883 & 5.27 & 27.76 &
38.6 & 33.1 \\
         & 2.44 & 3.21 & 100.45 & 94.46 & 88.25 & 48.32 & 0.935 & 38.98 &
47.55 & 63.4 & 52.1 \\
NGC 6093 & 1.36 & 2.79 & 5.04 & 3.73 & 4.71 & 3.56 & 0.149 & $-7.57$ & 23.62 &
43.3 & 47.8 \\
         & 0.87 & 1.53 & 3.30 & 2.95 & 3.23 & 2.55 & 0.327 & $-10.87$ & 0.53 &
29.6 & 31.5 \\
         & 3.41 & 4.05 & 6.47 & 5.75 & 6.00 & 5.02 & 0.174 & 24.31 & 32.75 &
53.3 & 58.9 \\
NGC 6121 & 0.43 & 0.92 & 6.86 & 5.81 & 4.55 & 1.75 & 0.728 & $-25.55$ &
$-10.73$ & 11.8 & 9.9 \\
         & 0.08 & 0.87 & 6.82 & 5.02 & 5.44 & 2.12 & 0.711 & $-18.80$ &
12.03 & 9.0 & 6.4 \\
         & 0.19 & 1.09 & 10.16 & 6.70 & 6.55 & 1.24 & 0.715 & $-44.32$ &
$-4.40$ & 16.5 & 15.2 \\
NGC 6144 & 1.95 & 2.23 & 3.13 & 2.88 & 2.95 & 2.60 & 0.127 & $-21.88$ &
$-15.18$ & 24.1 & 26.0 \\
         & 0.02 & 1.11 & 5.06 & 3.73 & 4.30 & 2.30 & 0.543 & $-22.46$ &
8.22 & 19.8 & 21.4 \\
         & 0.39 & 1.64 & 5.76 & 4.83 & 5.73 & 3.19 & 0.505 & $-21.05$ &
3.89 & 22.9 & 24.2 \\
NGC 6171 & 1.22 & 2.85 & 6.54 & 4.58 & 3.47 & 2.58 & 0.232 & 11.71 & 71.96 &
27.8 & 31.2 \\
         & 1.10 & 2.10 & 6.61 & 3.79 & 3.79 & 2.49 & 0.283 & 8.24 & 64.74 &
22.2 & 24.2 \\
         & 1.00 & 2.46 & 5.40 & 3.30 & 2.96 & 2.16 & 0.159 & 11.42 & 68.48 &
32.2 & 36.9 \\
NGC 6205 & 5.74 & 6.22 & 30.21 & 29.98 & 28.62 & 17.32 & 0.657 & $-61.63$ &
$-60.46$ & 80.9 & 79.3 \\
         & 4.24 & 4.89 & 18.85 & 18.34 & 17.90 & 11.53 & 0.580 & $-42.19$ &
$-38.96$ & 70.2 & 70.6 \\
         & 6.91 & 7.28 & 52.93 & 52.77 & 50.02 & 28.27 & 0.758 & $-87.90$ &
$-87.49$ & 89.0 & 83.3 \\
NGC 6218 & 1.71 & 2.53 & 10.32 & 6.19 & 4.20 & 2.66 & 0.417 & 33.79 & 85.64 &
31.5 & 34.1 \\
         & 0.14 & 1.12 &  6.97 & 4.37 & 3.20 & 2.00 & 0.608 & $-7.15$ & 48.77 &
23.9 & 24.3 \\
         & 2.78 & 3.38 & 10.38 & 8.19 & 4.78 & 3.30 & 0.409 & 60.03 & 91.64 &
33.6 & 36.2 \\
NGC 6254 & 0.75 & 3.07 & 7.12 & 5.55 & 3.40 & 2.47 & 0.291 & 18.09 & 80.59 &
38.3 & 42.6 \\
         & 0.73 & 1.92 & 6.97 & 4.81 & 3.24 & 1.78 & 0.430 & 16.42 & 70.69 &
34.2 & 37.1 \\
         & 3.73 & 4.31 & 9.32 & 7.29 & 4.27 & 3.22 & 0.251 & 76.27 & 100.81 &
43.5 & 47.7 \\
NGC 6266 & 0.62 & 1.44 & 3.11 & 2.50 & 1.15 & 0.89 & 0.278 & 15.94 & 49.27 &
42.1 & 48.6 \\
         & 0.62 & 1.16 & 2.30 & 1.89 & 1.21 & 0.98 & 0.250 & 10.61 & 35.26 &
34.6 & 39.5 \\
         & 1.32 & 2.87 & 5.46 & 3.81 & 1.47 & 1.15 & 0.147 & 27.96 & 87.40 &
58.3 & 66.4 \\
NGC 6304 & 1.42 & 3.40 & 7.02 & 5.14 & 1.04 & 0.75 & 0.203 & 32.96 & 105.75 &
31.9 & 35.9 \\
         & 1.03 & 1.74 & 2.95 & 2.58 & 0.68 & 0.61 & 0.205 & 23.86 & 55.16 &
23.8 & 27.6 \\
         & 1.63 & 2.38 & 5.06 & 4.24 & 0.68 & 0.59 & 0.287 & 40.70 & 77.09 &
29.8 & 33.0 \\
NGC 6316 & 0.13 & 0.78 & 5.07 & 3.37 & 2.98 & 0.89 & 0.616 & $-34.43$ & 18.64 &
20.3 & 20.8 \\
         & 0.02 & 0.49 & 2.84 & 1.96 & 1.69 & 0.81 & 0.610 & $-14.97$ & 20.38 &
15.0 & 13.6 \\
         & 1.58 & 1.75 & 5.00 & 4.64 & 2.17 & 1.44 & 0.451 & $-43.02$ &
$-38.34$ & 31.6 & 34.1 \\
NGC 6341 & 0.25 & 1.00 & 12.05 & 10.61 & 6.69 & 2.53 & 0.830 & 0.40 & 24.32 &
24.1 & 22.0 \\
         & 0.85 & 1.21 & 10.02 &  9.30 & 4.53 & 2.35 & 0.770 & 5.86 & 28.35 &
24.2 & 22.6 \\
         & 0.02 & 0.84 & 14.76 & 11.15 & 7.66 & 2.58 & 0.861 & $-14.16$ &
19.51 & 21.4 & 19.0 \\
NGC 6362 & 0.56 & 1.72 & 7.90 & 5.59 & 2.45 & 1.19 & 0.530 & 10.22 & 80.81 &
20.7 & 22.1 \\
         & 0.65 & 1.53 & 6.00 & 4.64 & 2.52 & 1.51 & 0.509 & 14.00 & 55.03 &
18.8 & 20.0 \\
         & 2.49 & 3.28 & 10.17 & 7.71 & 4.23 & 2.64 & 0.397 & 60.29 & 98.70 &
28.0 & 29.9 \\
NGC 6397 & 2.62 & 3.00 & 10.38 & 7.33 & 3.02 & 1.90 & 0.410 & 58.18 & 103.30 &
27.4 & 29.6 \\
         & 1.98 & 2.76 &  7.64 & 6.20 & 3.25 & 2.04 & 0.384 & 48.31 & 82.06 &
25.1 & 26.5 \\
         & 2.90 & 3.45 &  9.89 & 8.07 & 3.59 & 2.10 & 0.396 & 76.62 & 104.50 &
30.8 & 33.0 \\
NGC 6522 & 0.35 & 1.84 & 8.23 & 5.97 & 3.20 & 1.61 & 0.535 & 4.87 & 76.17 &
16.8 & 16.0 \\
         & 0.05 & 1.07 & 2.64 & 2.38 & 2.40 & 1.95 & 0.394 & $-7.47$ & 17.36 &
16.2 & 15.5 \\
         & 0.96 & 1.82 & 7.25 & 5.85 & 4.13 & 2.63 & 0.527 & 23.48 & 51.61 &
28.6 & 28.9 \\
NGC 6528 & 0.01 & 0.47 & 3.41 & 2.45 & 1.92 & 0.84 & 0.684 & $-16.74$ & 19.04 &
7.4 & 5.5 \\
         & 0.13 & 0.94 & 4.15 & 3.03 & 2.06 & 1.27 & 0.537 & $-7.01$ & 36.73 &
10.4 & 10.0 \\
         & 0.68 & 1.94 & 6.76 & 4.30 & 2.70 & 1.62 & 0.383 & 12.41 & 71.22 &
17.2 & 17.9 \\
NGC 6553 & 2.42 & 2.50 & 12.40 & 11.42 & 0.92 & 0.51 & 0.640 & 86.63 & 104.83 &
33.2 & 32.0 \\
         & 1.45 & 1.80 &  8.63 &  7.15 & 1.62 & 0.82 & 0.598 & 47.11 &  78.60 &
27.7 & 26.9 \\
         & 3.05 & 3.15 & 19.94 & 17.60 & 0.92 & 0.53 & 0.695 & 117.72 & 135.78 &
38.5 & 36.1 \\
NGC 6584 & 0.90 & 1.52 & 15.33 & 14.63 & 9.13 & 4.98 & 0.813 & 29.86 & 38.39 &
23.3 & 20.9 \\
         & 0.11 & 0.83 & 10.91 &  8.11 & 5.74 & 2.04 & 0.814 & $-8.08$ &
24.37 & 13.4 & 10.9 \\
         & 1.75 & 2.20 & 30.42 & 25.74 & 18.73 & 8.66 & 0.841 & 46.10 & 68.94 &
30.1 & 26.3 \\
NGC 6626 & 0.95 & 1.75 & 3.30 & 2.84 & 0.67 & 0.51 & 0.246 & 24.65 & 63.47 &
31.3 & 36.2 \\
         & 0.91 & 1.48 & 2.69 & 2.34 & 0.64 & 0.56 & 0.235 & 21.63 & 50.44 &
27.8 & 32.2 \\
         & 1.06 & 3.40 & 6.72 & 4.64 & 0.99 & 0.74 & 0.157 & 27.51 & 106.52 &
42.8 & 49.6 \\
NGC 6656 & 2.44 & 3.04 & 10.33 & 8.40 & 3.87 & 2.36 & 0.466 & 64.32 & 94.18 &
45.3 & 47.6 \\
         & 2.03 & 2.80 & 10.71 & 7.90 & 4.15 & 2.17 & 0.473 & 58.74 & 95.05 &
39.4 & 40.3 \\
         & 3.17 & 3.43 & 14.55 & 12.62 & 4.96 & 2.56 & 0.571 & 98.25 & 116.90 &
51.4 & 52.0 \\
NGC 6712 & 0.08 & 0.88 & 9.83 & 7.08 & 4.78 & 1.76 & 0.782 & $-1.26$ & 43.46 &
16.8 & 16.0 \\
         & 0.04 & 0.83 & 7.36 & 6.35 & 3.97 & 1.70 & 0.771 & $-8.50$ & 13.91 &
15.3 & 14.3 \\
         & 0.16 & 0.90 & 9.80 & 7.14 & 4.16 & 1.47 & 0.780 & $-2.57$ & 46.43 &
20.7 & 19.9 \\
NGC 6723 & 1.24 & 2.17 & 3.46 & 3.06 & 3.33 & 2.90 & 0.175 & $-0.97$ & 18.18 &
32.1 & 35.2 \\
         & 1.33 & 1.99 & 2.93 & 2.61 & 2.54 & 2.34 & 0.138 & 7.85 & 23.55 &
30.4 & 33.2 \\
         & 1.10 & 2.39 & 7.82 & 5.23 & 6.41 & 4.20 & 0.372 & $-15.42$ & 30.00 &
33.3 & 35.7 \\
NGC 6752 & 4.36 & 4.54 & 8.96 & 7.28 & 2.35 & 1.80 & 0.227 & 99.10 & 120.82 &
48.7 & 53.4 \\
         & 3.91 & 4.13 & 6.75 & 5.88 & 2.13 & 1.84 & 0.174 & 86.83 & 98.76 &
46.4 & 52.1 \\
         & 4.92 & 5.10 & 8.11 & 7.00 & 1.99 & 1.66 & 0.154 & 110.79 & 123.83 &
53.4 & 60.7 \\
NGC 6779 & 0.97 & 1.44 & 14.56 & 12.90 & 7.21 & 2.12 & 0.798 & $-65.00$ &
$-41.25$ & 17.2 & 14.6 \\
         & 0.15 & 0.73 & 14.96 & 12.19 & 7.84 & 1.34 & 0.888 & $-32.60$ &
$-5.74$ & 13.1 & 10.7 \\
         & 1.65 & 1.82 & 17.00 & 15.91 & 2.75 & 1.40 & 0.794 & $-84.26$ &
$-71.20$ & 23.1 & 20.5 \\
NGC 6809 & 0.11 & 1.77 & 7.73 & 6.03 & 6.33 & 4.21 & 0.556 & $-12.48$ & 25.49 &
23.9 & 23.3 \\
         & 0.39 & 1.80 & 11.86 & 6.82 & 6.28 & 3.67 & 0.578 & 7.20 & 61.32 &
30.5 & 32.0 \\
         & 0.07 & 1.14 &  8.55 & 7.45 & 7.80 & 4.45 & 0.746 & $-22.44$ & 7.10 &
17.3 & 15.4 \\
NGC 6838 & 4.66 & 4.69 & 8.42 & 7.74 & 0.35 & 0.30 & 0.244 & 121.05 & 131.50 &
26.6 & 30.0 \\
         & 4.01 & 4.07 & 9.23 & 8.18 & 0.41 & 0.32 & 0.333 & 109.22 & 126.49 &
24.3 & 26.5 \\
         & 5.28 & 5.39 & 7.90 & 7.24 & 0.38 & 0.34 & 0.146 & 129.07 & 137.68 &
29.3 & 33.8 \\
NGC 6934 & 6.91 & 7.21 & 56.21 & 55.73 & 54.84 & 31.27 & 0.771 & $-60.20$ &
$-58.98$ & 58.7 & 53.9 \\
         & 2.25 & 3.14 & 27.40 & 25.79 & 27.12 & 15.56 & 0.783 & $-22.81$ &
$-10.83$ & 33.5 & 30.3 \\
         & 9.95 & 10.08 & 211.67 & 211.41 & 200.73 & 115.01 & 0.909 & $-110.31$ &
$-110.19$ & 75.3 & 64.2 \\
NGC 7006 & 18.25 & 18.28 & 111.37 & 111.35 & 56.13 & 31.57 & 0.718 & 600.48 &
600.53 & 122.4 & 114.2 \\
         & 6.84 & 6.93 & 57.04 & 56.00 & 21.67 & 11.52 & 0.780 & 261.30 &
263.45 & 61.6 & 57.1 \\
       & \nodata & \nodata & \nodata & \nodata & \nodata & \nodata & \nodata &
\nodata & \nodata & \nodata & \nodata \\
NGC 7078 & 6.12 & 6.40 & 12.15 & 11.96 & 7.73 & 6.13 & 0.303 & 123.83 & 126.38 &
97.5 & 105.1 \\
         & 3.97 & 4.32 & 13.14 & 11.66 & 6.75 & 4.50 & 0.458 &  98.30 & 111.44 &
74.5 & 76.4 \\
         & 8.76 & 8.97 & 14.93 & 14.72 & 10.46 & 8.48 & 0.243 & 156.46 &
157.23 & 123.9 & 134.6 \\
NGC 7089 & 6.59 & 6.99 & 43.88 & 43.41 & 41.28 & 23.76 & 0.723 & $-76.05$ &
$-74.72$ & 96.5 & 89.9 \\
         & 5.60 & 6.10 & 23.72 & 23.17 & 23.15 & 14.65 & 0.583 & $-40.60$ &
$-38.26$ & 88.3 & 87.0 \\
         & 8.48 & 8.72 & 91.04 & 90.59 & 85.06 & 47.35 & 0.824 & $-127.98$ &
$-127.41$ & 112.1 & 98.1 \\
NGC 7099 & 3.63 & 3.94 & 7.87 & 7.58 & 6.25 & 4.95 & 0.316 & $-58.60$ &
$-55.62$ & 42.2 & 45.5 \\
         & 1.79 & 2.41 & 7.42 & 6.75 & 5.70 & 3.41 & 0.476 & $-50.57$ &
$-38.46$ & 29.7 & 30.1 \\
         & 5.00 & 5.42 & 9.38 & 9.16 & 7.76 & 6.39 & 0.257 & $-71.08$ &
$-69.88$ & 52.2 & 57.1 \\
Pal 12 & 15.45 & 15.60 & 26.24 & 26.15 & 23.46 & 18.71 & 0.253 & 172.48 &
172.55 & 44.0 & 47.5 \\
       & 13.60 & 13.78 & 15.57 & 15.41 & 14.05 & 13.35 & 0.056 & 122.29 &
122.39 & 40.9 & 45.7 \\
       & 17.19 & 17.30 & 46.11 & 46.05 & 40.87 & 27.77 & 0.454 & 235.90 &
235.95 & 46.3 & 47.3 \\
Pal 13 & 12.51 & 12.61 & 95.58 & 95.55 & 69.87 & 40.51 & 0.767 & $-339.98$ &
$-339.90$ & 28.1 & 25.9 \\
       &  9.73 &  9.90 & 57.77 & 57.72 & 43.74 & 25.04 & 0.707 & $-238.72$ &
$-238.57$ & 24.4 & 22.9 \\
       & 15.17 & 15.22 & 211.63 & 211.61 & 149.53 & 86.54 & 0.866 & $-457.13$ &
$-457.10$ & 32.0 & 28.3 \\
\enddata
\end{deluxetable}

\clearpage
\begin{deluxetable}{cccccccccc}
\tabletypesize{\scriptsize}
\rotate
\tablecaption{Destruction Rates \label{tbl-3}}
\tablewidth{0pt}
\tablehead{
\colhead{Cluster} & \colhead{$M_c$} & \colhead{$c$} & \colhead{$r_h$} &
\colhead{$<$$1/t_{bulge,1}$$>$} &
\colhead{$<$$1/t_{bulge,2}$$>$} & \colhead{$<$$1/t_{bulge}$$>$} &
\colhead{$<$$1/t_{disk,1}$$>$} & \colhead{$<$$1/t_{disk,2}$$>$} &
\colhead{$<$$1/t_{disk}$$>$} \\ \colhead{} & \colhead{($M_{\odot}$)} &
\colhead{} & \colhead{($pc$)} &
\colhead{($yr^{-1}$)} & \colhead{($yr^{-1}$)} & \colhead{($yr^{-1}$)} & 
\colhead{($yr^{-1}$)} & \colhead{($yr^{-1}$)} & \colhead{($yr^{-1}$)}
}
\startdata
NGC 104 & 1.0E06 & 2.03 & 3.65 & 1.1E-16 & 6.4E-18 & 1.1E-16 & 8.8E-14 &
9.8E-15 & 9.8E-14 \\
 &  &  &  & 1.3E-16 & 7.5E-18 & 1.3E-16 & 8.7E-14 & 9.6E-15 & 9.7E-14 \\
NGC 288 & 8.5E04 & 0.96 & 5.68 & 1.9E-13 & 5.7E-14 & 2.5E-13 & 1.8E-12 &
8.5E-13 & 2.7E-12 \\
 &  &  &  & 2.2E-13 & 6.4E-14 & 2.8E-13 & 1.7E-12 & 8.2E-13 & 2.6E-12 \\
NGC 362 & 3.9E05 & 1.94 & 2.00 & 4.4E-11 & 6.4E-12 & 5.0E-11 & 7.7E-13 &
8.5E-14 & 8.6E-13 \\
 &  &  &  & 8.1E-12 & 1.1E-12 & 9.3E-12 & 4.8E-13 & 5.9E-14 & 5.3E-13 \\
NGC 1851 & 3.7E05 & 2.32 & 1.83 & 1.5E-16 & 7.6E-18 & 1.5E-16 & 6.9E-15 &
5.8E-16 & 7.5E-15 \\
 &  &  &  & 1.5E-16 & 7.6E-18 & 1.6E-16 & 6.8E-15 & 5.7E-16 & 7.4E-15 \\
NGC 1904 & 2.4E05 & 1.72 & 3.00 & 1.3E-15 & 1.5E-16 & 1.4E-15 & 4.2E-14 &
8.0E-15 & 5.0E-14 \\ 
 &  &  &  & 1.5E-15 & 1.8E-16 & 1.7E-15 & 5.0E-14 & 1.0E-14 & 5.9E-14 \\
NGC 2298 & 5.7E04 & 1.28 & 2.43 & 1.8E-15 & 2.6E-16 & 2.0E-15 & 9.1E-14 &
2.5E-14 & 1.2E-13 \\
 &  &  &  & 1.8E-15 & 2.6E-16 & 2.0E-15 & 8.8E-14 & 2.4E-14 & 1.1E-13 \\
NGC 4147 & 5.0E04 & 1.80 & 2.41 & 9.0E-13 & 1.5E-13 & 1.1E-12 & 4.5E-13 &
9.7E-14 & 5.5E-13 \\
 &  &  &  & 9.5E-13 & 1.6E-13 & 1.1E-12 & 5.8E-13 & 1.3E-13 & 7.1E-13 \\
NGC 4590 & 1.5E05 & 1.64 & 4.60 & 6.3E-15 & 5.8E-16 & 6.9E-15 & 1.1E-13 &
1.5E-14 & 1.2E-13 \\
 &  &  &  & 6.4E-15 & 5.9E-16 & 7.0E-15 & 1.1E-13 & 1.5E-14 & 1.2E-13 \\
NGC 5024 & 5.2E05 & 1.78 & 5.75 & 5.7E-17 & 3.5E-18 & 6.1E-17 & 7.4E-16 &
6.6E-17 & 8.1E-16 \\
 &  &  &  & 5.7E-17 & 3.6E-18 & 6.1E-17 & 7.4E-16 & 6.6E-17 & 8.1E-16 \\
NGC 5139 & 3.3E06 & 1.61 & 6.44 & 5.7E-12 & 8.8E-13 & 6.6E-12 & 2.7E-12 &
4.2E-13 & 3.1E-12 \\
 &  &  &  & 2.3E-12 & 3.3E-13 & 2.6E-12 & 2.2E-12 & 3.6E-13 & 2.5E-12 \\ 
NGC 5272 & 6.4E05 & 1.84 & 3.39 & 1.5E-13 & 9.1E-15 & 1.6E-13 & 9.4E-13 &
8.7E-14 & 1.0E-12 \\
 &  &  &  & 1.5E-13 & 9.0E-15 & 1.6E-13 & 8.5E-13 & 8.0E-14 & 9.3E-13 \\ 
NGC 5466 & 1.0E05 & 1.32 & 10.41 & 1.1E-12 & 2.1E-13 & 1.3E-12 & 2.4E-12 &
5.9E-13 & 2.9E-12 \\ 
 &  &  &  & 1.1E-12 & 2.2E-13 & 1.3E-12 & 2.4E-12 & 5.8E-13 & 2.9E-12 \\
Pal 5 & 2.0E04 & 0.70 & 19.97 & 9.3E-11 & 4.9E-11 & 1.4E-10 & 2.6E-10 &
1.5E-10 & 4.1E-10 \\
 &  &  &  & 9.3E-11 & 4.9E-11 & 1.4E-10 & 2.5E-10 & 1.5E-10 & 4.1E-10 \\
NGC 5897 & 1.3E05 & 0.79 & 7.61 & 1.7E-10 & 8.6E-11 & 2.6E-10 & 1.5E-11 &
7.6E-12 & 2.3E-11 \\
 &  &  &  & 3.6E-11 & 1.6E-11 & 5.2E-11 & 1.5E-11 & 7.5E-12 & 2.3E-11 \\
NGC 5904 & 5.7E05 & 1.83 & 4.60 & 5.6E-13 & 1.0E-13 & 6.6E-13 & 1.6E-13 &
3.6E-14 & 1.9E-13 \\
 &  &  &  & 4.8E-13 & 8.7E-14 & 5.7E-13 & 1.5E-13 & 3.3E-14 & 1.8E-13 \\
NGC 6093 & 3.3E05 & 1.95 & 1.89 & 3.9E-14 & 2.6E-15 & 4.1E-14 & 1.1E-12 &
1.4E-13 & 1.3E-12 \\
 &  &  &  & 1.6E-13 & 1.3E-14 & 1.8E-13 & 1.1E-12 & 1.3E-13 & 1.2E-12 \\
NGC 6121 & 1.3E05 & 1.59 & 2.33 & 1.8E-10 & 5.7E-11 & 2.3E-10 & 7.4E-13 &
1.4E-13 & 8.9E-13 \\
 &  &  &  & 2.9E-11 & 8.3E-12 & 3.8E-11 & 6.8E-13 & 1.2E-13 & 8.0E-13 \\
NGC 6144 & 8.6E04 & 1.55 & 4.00 & 1.4E-10 & 1.9E-11 & 1.6E-10 & 2.8E-10 &
5.0E-11 & 3.3E-10 \\
 &  &  &  & 1.6E-10 & 2.2E-11 & 1.8E-10 & 2.3E-10 & 4.0E-11 & 2.7E-10 \\
NGC 6171 & 1.2E05 & 1.51 & 5.03 & 2.4E-12 & 6.6E-13 & 3.0E-12 & 8.9E-12 &
3.4E-12 & 1.2E-11 \\
 &  &  &  & 5.7E-12 & 1.8E-12 & 7.5E-12 & 8.4E-12 & 3.2E-12 & 1.2E-11 \\
NGC 6205 & 5.2E05 & 1.51 & 3.34 & 5.5E-16 & 4.3E-17 & 5.9E-16 & 2.5E-14 &
3.9E-15 & 2.9E-14 \\
 &  &  &  & 5.5E-16 & 4.3E-17 & 5.9E-16 & 2.5E-14 & 3.9E-15 & 2.9E-14 \\
NGC 6218 & 1.4E05 & 1.39 & 3.08 & 2.2E-13 & 4.1E-14 & 2.6E-13 & 9.9E-13 &
2.6E-13 & 1.3E-12 \\
 &  &  &  & 1.2E-13 & 2.2E-14 & 1.4E-13 & 9.5E-13 & 2.5E-13 & 1.2E-12 \\
NGC 6254 & 1.7E05 & 1.40 & 2.32 & 1.0E-14 & 1.0E-15 & 1.1E-14 & 5.8E-13 &
9.8E-14 & 6.8E-13 \\
 &  &  &  & 7.8E-15 & 7.8E-16 & 8.6E-15 & 5.3E-13 & 8.9E-14 & 6.2E-13 \\
NGC 6266 & 8.1E05 & 1.70 & 2.47 & 1.6E-14 & 2.0E-15 & 1.8E-14 & 1.1E-14 &
1.5E-15 & 1.2E-14 \\
 &  &  &  & 3.3E-15 & 4.0E-16 & 3.7E-15 & 1.0E-14 & 1.4E-15 & 1.1E-14 \\
NGC 6304 & 1.5E05 & 1.80 & 2.46 & 6.5E-13 & 1.0E-13 & 7.5E-13 & 1.3E-12 &
2.0E-13 & 1.5E-12 \\
 &  &  &  & 1.5E-13 & 2.1E-14 & 1.7E-13 & 7.4E-13 & 1.1E-13 & 8.4E-13 \\
NGC 6316 & 3.7E05 & 1.55 & 2.27 & 1.8E-12 & 3.3E-13 & 2.1E-12 & 1.0E-13 &
1.6E-14 & 1.2E-13 \\
 &  &  &  & 1.5E-12 & 3.0E-13 & 1.8E-12 & 8.5E-14 & 1.4E-14 & 9.8E-14 \\
NGC 6341 & 3.3E05 & 1.81 & 2.60 & 5.8E-12 & 9.4E-13 & 6.8E-12 & 7.3E-13 &
1.1E-13 & 8.4E-13 \\
 &  &  &  & 3.2E-12 & 5.2E-13 & 3.8E-12 & 5.9E-13 & 8.5E-14 & 6.7E-13 \\
NGC 6362 & 1.0E05 & 1.10 & 4.82 & 6.7E-12 & 1.8E-12 & 8.5E-12 & 2.2E-11 &
6.9E-12 & 2.9E-11 \\
 &  &  &  & 4.0E-11 & 1.2E-11 & 5.2E-11 & 3.8E-11 & 1.2E-11 & 5.0E-11 \\
NGC 6397 & 7.7E04 & 2.50 & 1.56 & 3.0E-16 & 3.3E-17 & 3.3E-16 & 1.1E-14 &
1.9E-15 & 1.3E-14 \\
 &  &  &  & 2.9E-16 & 3.2E-17 & 3.2E-16 & 1.2E-14 & 2.1E-15 & 1.4E-14 \\ 
NGC 6522 & 2.0E05 & 2.50 & 2.36 & 3.1E-10 & 5.2E-11 & 3.6E-10 & 1.4E-11 &
2.2E-12 & 1.6E-11 \\
 &  &  &  & 2.1E-10 & 3.9E-11 & 2.5E-10 & 1.1E-11 & 1.7E-12 & 1.3E-11 \\
NGC 6528 & 7.2E04 & 2.29 & 0.99 & 2.2E-09 & 1.8E-10 & 2.4E-09 & 7.8E-11 &
5.0E-12 & 8.3E-11 \\
 &  &  &  & 3.4E-09 & 3.2E-10 & 3.7E-09 & 8.8E-11 & 4.6E-12 & 9.3E-11 \\
NGC 6553 & 2.2E05 & 1.17 & 2.70 & 8.1E-16 & 1.5E-16 & 9.6E-16 & 2.4E-15 &
3.4E-16 & 2.8E-15 \\
 &  &  &  & 9.8E-16 & 1.7E-16 & 1.2E-15 & 3.2E-15 & 4.6E-16 & 3.7E-15 \\
NGC 6584 & 2.0E05 & 1.20 & 3.12 & 4.4E-11 & 1.1E-11 & 5.5E-11 & 1.3E-12 &
2.9E-13 & 1.6E-12 \\
 &  &  &  & 7.0E-12 & 1.6E-12 & 8.6E-12 & 1.1E-12 & 2.7E-13 & 1.4E-12 \\
NGC 6626 & 3.2E05 & 1.67 & 2.54 & 5.1E-15 & 6.4E-16 & 5.7E-15 & 5.0E-14 &
7.1E-15 & 5.8E-14 \\ 
 &  &  &  & 5.3E-14 & 8.6E-15 & 6.1E-14 & 7.8E-14 & 1.2E-14 & 8.9E-14 \\
NGC 6656 & 4.3E05 & 1.31 & 3.03 & 1.9E-15 & 2.3E-16 & 2.1E-15 & 2.8E-14 &
4.6E-15 & 3.3E-14 \\
 &  &  &  & 5.4E-15 & 6.8E-16 & 6.1E-15 & 7.1E-14 & 1.3E-14 & 8.4E-14 \\
NGC 6712 & 1.7E05 & 0.90 & 2.75 & 2.5E-12 & 7.8E-13 & 3.3E-12 & 1.3E-13 &
3.5E-14 & 1.6E-13 \\
 &  &  &  & 1.3E-12 & 4.1E-13 & 1.7E-12 & 1.1E-13 & 3.1E-14 & 1.5E-13 \\
NGC 6723 & 2.3E05 & 1.05 & 4.07 & 2.8E-13 & 5.8E-14 & 3.3E-13 & 1.8E-12 &
5.9E-13 & 2.3E-12 \\
 &  &  &  & 6.0E-13 & 1.4E-13 & 7.4E-13 & 1.6E-12 & 5.5E-13 & 2.2E-12 \\ 
NGC 6752 & 2.1E05 & 2.50 & 2.72 & 3.3E-14 & 2.1E-15 & 3.5E-14 & 1.6E-11 &
1.6E-12 & 1.7E-11 \\
 &  &  &  & 1.4E-13 & 1.0E-14 & 1.5E-13 & 1.1E-11 & 1.1E-12 & 1.2E-11 \\ 
NGC 6779 & 1.5E05 & 1.37 & 3.41 & 8.0E-12 & 2.5E-12 & 1.1E-11 & 7.0E-13 &
1.5E-13 & 8.5E-13 \\
 &  &  &  & 4.0E-12 & 1.2E-12 & 5.2E-12 & 7.1E-13 & 1.5E-13 & 8.6E-13 \\
NGC 6809 & 1.8E05 & 0.76 & 4.45 & 1.3E-11 & 4.9E-12 & 1.8E-11 & 1.6E-12 &
6.7E-13 & 2.3E-12 \\ 
 &  &  &  & 6.3E-12 & 2.3E-12 & 8.6E-12 & 1.4E-12 & 6.0E-13 & 2.0E-12 \\
NGC 6838 & 3.0E04 & 1.15 & 1.92 & 8.6E-17 & 1.2E-17 & 9.8E-17 & 1.5E-14 &
1.7E-15 & 1.6E-14 \\
 &  &  &  & 9.3E-17 & 1.3E-17 & 1.1E-16 & 1.2E-14 & 1.4E-15 & 1.3E-14 \\
NGC 6934 & 1.6E05 & 1.53 & 2.74 & 1.4E-16 & 1.3E-17 & 1.6E-16 & 8.1E-15 &
1.5E-15 & 9.6E-15 \\ 
 &  &  &  & 1.4E-16 & 1.3E-17 & 1.6E-16 & 8.1E-15 & 1.5E-15 & 9.6E-15 \\
NGC 7006 & 2.0E05 & 1.42 & 4.59 & 7.9E-18 & 5.8E-19 & 8.5E-18 & 4.9E-17 &
3.9E-18 & 5.2E-17 \\
 &  &  &  & 7.9E-18 & 5.9E-19 & 8.5E-18 & 4.9E-17 & 3.9E-18 & 5.2E-17 \\
NGC 7078 & 8.0E05 & 2.50 & 3.17 & 5.6E-16 & 3.0E-17 & 5.9E-16 & 9.1E-14 &
9.5E-15 & 1.0E-13 \\
 &  &  &  & 5.6E-16 & 3.0E-17 & 5.9E-16 & 8.9E-14 & 9.2E-15 & 9.8E-14 \\
NGC 7089 & 6.9E05 & 1.80 & 3.11 & 3.3E-16 & 2.0E-17 & 3.5E-16 & 1.1E-14 &
1.3E-15 & 1.2E-14 \\
 &  &  &  & 3.4E-16 & 2.0E-17 & 3.6E-16 & 1.1E-14 & 1.3E-15 & 1.2E-14 \\
NGC 7099 & 1.6E05 & 2.50 & 2.68 & 5.4E-14 & 5.0E-15 & 5.9E-14 & 2.1E-12 &
3.4E-13 & 2.4E-12 \\ 
 &  &  &  & 6.1E-14 & 5.7E-15 & 6.7E-14 & 2.0E-12 & 3.3E-13 & 2.4E-12 \\
Pal 12 & 1.1E04 & 1.94 & 7.11 & 1.1E-13 & 1.9E-14 & 1.3E-13 & 1.1E-12 &
2.2E-13 & 1.3E-12 \\
 &  &  &  & 1.1E-13 & 1.9E-14 & 1.3E-13 & 1.1E-12 & 2.2E-13 & 1.3E-12 \\
Pal 13 & 5.4E03 & 0.68 & 3.45 & 9.5E-16 & 2.2E-16 & 1.2E-15 & 1.1E-14 &
3.6E-15 & 1.5E-14 \\
 &  &  &  & 9.5E-16 & 2.2E-16 & 1.2E-15 & 1.1E-14 & 3.6E-15 & 1.5E-14 \\
\enddata
\end{deluxetable}

\clearpage

\clearpage

\clearpage

\clearpage

\clearpage

\clearpage

\clearpage

\clearpage

\clearpage

\clearpage

\clearpage

\clearpage

\end{document}